\documentclass[11pt,a4paper,onecolumn,reqno]{amsart}
\usepackage[a4paper, margin=2.3cm, bmargin=3cm]{geometry}

\usepackage{graphicx,stfloats} \usepackage{color}
\usepackage{amsmath}
\usepackage{amsaddr}
\usepackage{bm, soul}
\usepackage{mathtools}
\usepackage[colorinlistoftodos,prependcaption,textsize=tiny]{todonotes}
\usepackage{braket}
\usepackage{hyperref}
\usepackage[square,comma,sort&compress, numbers]{natbib}
\usepackage{enumerate}
\usepackage[version=4]{mhchem}

\usepackage{tikz}
\usepackage{pgfplots}
\pgfplotsset{compat=1.9}
\usetikzlibrary{calc,arrows,fadings,decorations.pathreplacing,decorations.markings,patterns,shapes.geometric}
\tikzset{>=stealth,inner sep=0pt, outer sep=2pt,}
\tikzset{vecteur/.style={->,thick,color=black,smooth}}

\newcommand{\reviewtwo}{\textcolor{black}}

\usepackage{placeins}

\renewcommand{\st}[1]{}

\usepackage{siunitx}
 
\usepackage{multirow}

\usepackage{etoolbox}
\patchcmd{\pprintMaketitle}
  {\hrule\vskip12pt}
 {\hrule\vskip12pt\ifvoid\extrainfobox\else\unvbox\extrainfobox\par\vskip12pt\fi}
 {}{}

\newsavebox\extrainfobox

\usepackage{caption}
\title{Turbulent flame-wall interaction of premixed flames using Quadrature-based Moment Methods (QbMM) and tabulated chemistry: an \textit{\textit{a priori}} analysis}

\author[stfs]{Matthias Steinhausen$^{1,*}$, Thorsten Zirwes$^{2,3}$, Federica Ferraro$^1$, Sebastian Popp$^{1}$, Feichi Zhang$^{2}$, Henning Bockhorn$^{2}$, Christian Hasse$^{1}$}
\email{steinhausen@stfs.tu-darmstadt.de} 
\address[]{$^1$Technical University of Darmstadt, Department of Mechanical Engineering, Simulation of reactive Thermo-Fluid Systems, Otto-Berndt-Stra{\ss}e 2, 64287 Darmstadt, Germany\\
$^2$Engler-Bunte-Institute, Karlsruhe Institute of Technology, Engler-Bunte-Ring 7, 76131 Karlsruhe, Germany\\
$^3$Steinbuch Centre for Computing, Karlsruhe Institute of Technology, Hermann-von-Helmholtz-Platz 1, 76344 Eggenstein-Leopoldshafen, Germany
}

\begin{document}
\pagestyle{plain}

\maketitle
\begin{abstract}
Presumed probability density function and transported PDF methods are commonly applied to model the turbulence chemistry interaction in turbulent reacting flows. However, little focus has been given to the turbulence chemistry interaction PDF closure for flame-wall interaction. In this study, a quasi-DNS of a turbulent premixed, stoichiometric methane-air flame ignited in a fully developed turbulent channel flow undergoing side-wall quenching is investigated. The objective of this study is twofold. First, the joint PDF of the progress variable and enthalpy, that needs to be accounted for in turbulence chemistry interaction closure models, is analyzed in the quasi-DNS configuration, both in the core flow and the near-wall region. Secondly, a transported PDF closure model, based on a Conditional Quadrature Method of Moments approach, and a presumed PDF approach are examined in an \textit{a priori} analysis using the quasi-DNS as a reference both in the context of Reynolds-Averaged Navier Stokes (RANS) and Large-Eddy Simulations (LESs).
The analysis of the joint PDF demonstrates the high complexity of the reactive scalar distribution in the near-wall region. Here, a high correlation between the progress variable and enthalpy is found, where the flame propagation and quenching are present simultaneously.
The transported PDF approach presented in this work, based on the Conditional Quadrature Method of Moments, accounts for the moments of the joint PDF of progress variable and enthalpy coupled to a Quenching-Flamelet Generated Manifold. In the \textit{a priori} analysis  both turbulence chemistry interaction PDF closure models show a high accuracy in the core flow. In the near-wall region, however, only the Conditional Quadrature Method of Moments approach is suitable to predict the flame structures.
\end{abstract}

\keywords{\textbf{Keywords:} Quadrature-based Moment Methods; Flame-wall interaction; Turbulent premixed flame; Direct Numerical Simulation (DNS); Turbulent side-wall quenching}

\section{Introduction}

In most industrial systems the combustion takes place in a vessel to allow the generation of power. In the vessel, flames develop in the vicinity of walls and interact with them, leading to flame-wall interactions (FWIs) that lower the overall combustion efficiency, impact the pollutant formation~\cite{Poinsot2005a} and can also lead to undesired flame behavior, such as flame flashback~\cite{Fritz2001}. In turbulent FWIs, turbulence increases the complexity even further.

Numerical simulations of turbulent flames in the close vicinity of walls pose two major challenges. First, the thermochemical reactions inside the flame that are influenced by heat losses to the (cold) walls need to be considered and, secondly, it is necessary to account for the fluctuations of the reactive scalars caused by the turbulence. In direct numerical simulations (DNSs) of turbulent FWI~\cite{Gruber2010, Ahmed2021}, typically, finite-rate chemistry is used to model the reactions in the flame, while the fluctuations of the reactive scalars are resolved by the simulation. This, however, leads to high computational costs and is not suitable for real combustion applications. Here, Reynolds-Averaged Navier Stokes (RANS) or Large-Eddy Simulations (LESs) are employed that use reduced or tabulated chemistry approaches and require turbulence chemistry interaction (TCI) closure models to account for the unresolved fluctuations of the reactive scalars. In the context of tabulated chemistry approaches, chemistry manifolds for FWI have been studied extensively~\cite{VanOijen2000, Ganter2017, Ganter2018, Strassacker2019a, Efimov2019b, Steinhausen2021} using mostly experiments of premixed, laminar methane-air flames~\cite{Jainski2017, Jainski2017a, Kosaka2018, Kosaka2019} as a reference. These manifolds have also been applied for the simulation of turbulent flames using RANS~\cite{Fiorina2005} and LES~\cite{Heinrich2018a, Heinrich2018b, Wu2015}.

For the TCI closure, multiple approaches are reported in the literature. In the context of this work, TCI closure models for the joint probability density function (PDF) are discussed. In these PDF based methods the unresolved fluctuations of the reactive scalars are modeled by their statistical behavior that is captured in a PDF in the context of RANS and a filtered density function (FDF) in the context of LES. In presumed PDF (pPDF) approaches, the PDF shape is assumed and parameterized by low order moments, e.g. the mean and the variance. They have been applied in multiple studies of turbulent premixed flames~\cite{Fiorina2005, Bray2006, Jin2008, Fiorina2010a, Salehi2013a, Donini2017,Zhang2021} also considering enthalpy losses in the flame using a $\beta$-PDF for the progress variable and a $\delta$-peak for the enthalpy~\cite{Fiorina2005,Donini2017,Zhang2021}. However, it is only in~\cite{Fiorina2005} that the focus is particularly on FWI. In transported PDF (tPDF) models the whole (joint) PDF is solved for during runtime and the statistical behavior of the unresolved fluctuations is represented by a one-point, one-time joint PDF of relevant flow variables~\cite{Muradoglu1999}. Different Monte-Carlo methods were established to solve the high-dimensional PDF efficiently. The Lagrangian approach~\cite{Pope1981} models the PDF transport equation by the evolution of a large set of stochastic particles, while the Eulerian stochastic fields (SF) approach~\cite{Valino1998} solves for a set of stochastic fields. Both Lagrangian and Eulerian PDF methods are constructed such that their one-point one-time joint PDF corresponds to that of a real turbulent reacting system~\cite{Haworth2010}. Using these methods, non-premixed~\cite{Jones2010,Cao2005,Ferraro2019}, partially premixed~\cite{Jones2009} and premixed flames~\cite{Avdic2016, Tirunagari2016} have been simulated.

A computationally more efficient approach to solve PDF-based systems is the Method of Moments. In contrast to the Monte-Carlo methods, a set of integral PDF properties, i.e., its moments, are solved. The Method of Moments has been successfully applied to various applications, including nano-particles and aerosols~\cite{Mcgraw1997}, sprays~\cite{Pollack2016} and combustion-related problems, such as soot formation~\cite{Salenbauch2019, Wick2017, Ferraro2021}. For the closure of the joint PDF, similarly to Monte-Carlo transported PDF methods, the Method of Moments was primarily applied to turbulent non-premixed flames~\cite{Raman2006, Tang2007, Koo2011, Jaishree2012, Donde2012, MadadiKandjani2017}. In~\cite{Pollack2021}, the Quadrature Method of Moments (QMOM)~\cite{Mcgraw1997} was applied to premixed turbulent flames. In the study, the thermochemical state $\phi$ is reconstructed from a chemistry manifold $f$ as a function of the progress variable $Y_c$, i.e. $\phi = f\left( Y_\mathrm{c} \right)$. Closure of the joint PDF reduces to a  univariate representation of the PDF, in this case for the progress variable $Y_c$. The QMOM approach showed similar accuracy to previous Monte-Carlo simulations while reducing the computational costs.

As outlined above, TCI in premixed flames has been studied extensively, however, the closure of the joint PDF for FWI has received little attention. This work focuses on the analysis and modeling of the joint PDF of the progress variable and enthalpy in the context of turbulent FWI. A quasi-DNS of a stoichiometric methane-air flame ignited in a fully developed turbulent channel flow is performed. Note that the term quasi-DNS~\cite{Zirwes2020} indicates that all scales in the region of interest are resolved (Kolmogorov length, flame thickness, $y^+$), but the convergence-order of the numerical schemes are limited to fourth-order for spatial schemes and second-order for the time discretization. The setup is inspired by the DNS of a hydrogen-air flame by Gruber et al.~\cite{Gruber2010}. Using the quasi-DNS data, both the marginal PDFs/FDFs of the progress variable and enthalpy and their joint PDF/FDF are analyzed in the context of RANS/LES and the unresolved fluctuations in the near-wall region are taken into particular consideration. The open question posed in~\cite{Fiorina2005} of the statistical independence of the progress variable and enthalpy in the close vicinity of the wall is addressed, giving useful insights for closure models of the joint PDF in the context of FWI. Based on these insights, a novel tPDF approach is presented and compared to a pPDF approach from the literature~\cite{Fiorina2005,Donini2017,Zhang2021}. The tPDF closure model is an extension of the work of Pollack et al.~\cite{Pollack2021}. It couples the joint PDF of the progress variable and enthalpy using a Conditional QMOM (CQMOM) approach~\cite{Cheng2010} with a Quenching-Flamelet Generated Manifold (QFM)~\cite{Efimov2019b}. The suitability of the CQMOM closure is assessed by means of an \textit{a priori} analysis~\cite{Bray2006, Jin2008}.

In the first part of this work, the quasi-DNS of the generic side-wall quenching configuration is introduced. Then, the chemistry manifold employed in this work is presented together with the novel CQMOM approach for the closure of the joint PDF. Using the quasi-DNS data the PDFs/FDFs of the progress variable and enthalpy are analyzed in the context of RANS/LES. In the second part, the CQMOM approach is evaluated in an \textit{a priori} analysis with a special focus on the near-wall region and how it differs from the core region with unconstrained flame propagation.


\section{Quasi-DNS of a turbulent side-wall quenching configuration}
\begin{figure}[htb]
    \centering
    \begin{tikzpicture} [scale=8.5/11]

        \draw[thick](0,-1)--++(10,0);
    	\fill[pattern=north west lines] (0,-1.3) rectangle (10,-1);
    	\draw[thick](0,1)--++(10,0);
    	\fill[pattern=north west lines] (0,1) rectangle (10,1.3);
        \draw[lightgray, dash dot] (-0.1,0)--++(10.2,0) node[right]{};

    	\foreach \y in {-0.9,-0.7,-0.5,-0.3,-0.1,0.9,0.7,0.5,0.3,0.1}
    	\draw[vecteur] (0,\y)--++({2*(1-(\y*\y)^(1/2))^(1/7)-1.2},0);
        \draw (1,0.7)node{\scriptsize $U_\mathrm{bulk}$};

        \draw[|<->|] (0,1.8)--++(10,0) node[midway, above]{\scriptsize $10H$};
        \draw[|<->|] (8,-1.0)--++(0,1) node[midway, right]{\scriptsize $H$};

        \draw[ultra thick, red] plot [smooth] coordinates { (9,0.9) (7, 0.78) (6, 0.65) (4.5, 0.4) (3, 0.1) (1.4, -0.25) (0.95, -0.5) (1.4, -0.7) (5, -0.9) (5.8, -0.95) (6, -1) };
        \node[anchor=center, red] at (5.5, -0.6) {\scriptsize lower flame branch};
        \node[red, inner sep=2pt, fill=white, anchor=center] at (5.5, 0.6) {\scriptsize upper flame branch};

    	\filldraw (1,-0.5) circle (0.09);
        \draw[-] (1, -0.5) -- (1.3,-0.45) node[right]{\scriptsize flame holder};
        \draw[|<->|] (-0.3,-1)--++(0,0.5) node[midway, left]{\scriptsize $H/2$};
        \draw[|<->|] (0,-1.8)--++(1,0) node[midway, below]{\scriptsize $H$};

        \draw[->] (9.5,-2.0)--++(0,0.5) node[above, right]{\scriptsize $z$};
        \draw[->] (9.5,-2.0)--++(0.5,0) node[above, right]{\scriptsize $x$};
        \node[anchor=center,draw,circle] at (9.5, -2.0) {\tiny \textbullet};
        \node[anchor=center] at (9.2, -2.0) {\scriptsize $y$};

        \node[gray, inner sep=2pt, fill=white, anchor=center,rotate=270] at (9.8, 0.0) {\footnotesize outlet};
        \node[gray, inner sep=2pt, fill=white, anchor=center,rotate=270] at (0.2, 0.0) {\footnotesize inlet};
        \node[gray, inner sep=2pt, fill=white, anchor=center] at (5, -1.33) {\footnotesize isothermal wall ($T=300$ K)};
        \node[gray, inner sep=2pt, fill=white, anchor=center] at (5, 1.33) {\footnotesize isothermal wall ($T=300$ K)};

    \end{tikzpicture}
   \caption{2D sketch of the reactive quasi-DNS. The flame is depicted in red and the boundaries of the domain are labeled in gray. In the lateral direction ($y$) the channel width is $3H$ and the flow is assumed to be statistically independent, hence, periodic boundary conditions are applied.}
  \label{fig:DNS-setup}
\end{figure}
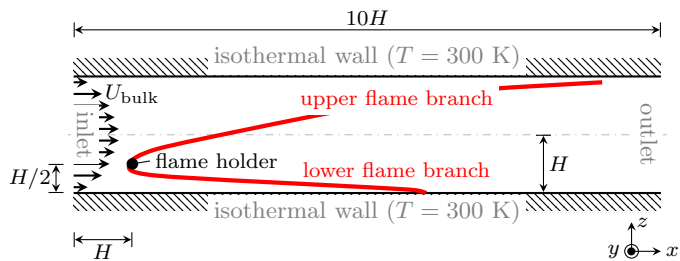

The generic side-wall quenching case analyzed in this work is inspired by~\cite{Gruber2010}. A V-shaped, premixed stoichiometric methane-air flame anchored in a fully developed turbulent channel flow is simulated. Figure~\ref{fig:DNS-setup} shows a sketch of the quasi-DNS setup. At the inlet the unburnt stochiometric methane-air mixture at $T=300~\text{K}$ enters the channel with an inflow velocity corresponding to a fully developed channel flow at a flow Reynolds number of $\mathrm{Re} = \left( U_\text{bulk} H \right) / \nu \approx 2770$, with $U_\text{bulk}$ being the mean flow velocity, $H$ the channel half-width and $\nu$ the dynamic viscosity. This corresponds to $\mathrm{Re}_\tau = H/\delta_v = 180$, with $\delta_v=\nu \cdot \sqrt{\rho / \tau_{w}}$ being the viscous length scale, $\rho$ the density of the unburnt methane-air mixture and $\tau_{w}$ the wall shear stress. The mean inflow velocity is chosen to be $U_\text{bulk}=4.4~\mathrm{ms^{-1}}$, leading to a channel half-width of $H=10~\mathrm{mm}$. The flame is anchored at $H/2$ above the bottom wall inside the core flow of the channel. The flame holder is not modeled as a physical object (i.e. a wall-boundary in the simulation), but instead simply as a cylindrical region ($r=0.09\cdot H$) of burnt gas temperature. Because of this, the influence of the numerical flame holder on the flow field is given by thermal expansion and thus acceleration of the flow. A similar flame anchor has also been used in previous studies~\cite{Gruber2010}. At the flame holder a V-shaped flame is formed that propagates through the turbulent boundary layer to the isothermal channel walls ($T_\mathrm{wall}=300~\text{K}$), where it finally quenches.

\begin{figure}[htb]
  \includegraphics[width=\textwidth]{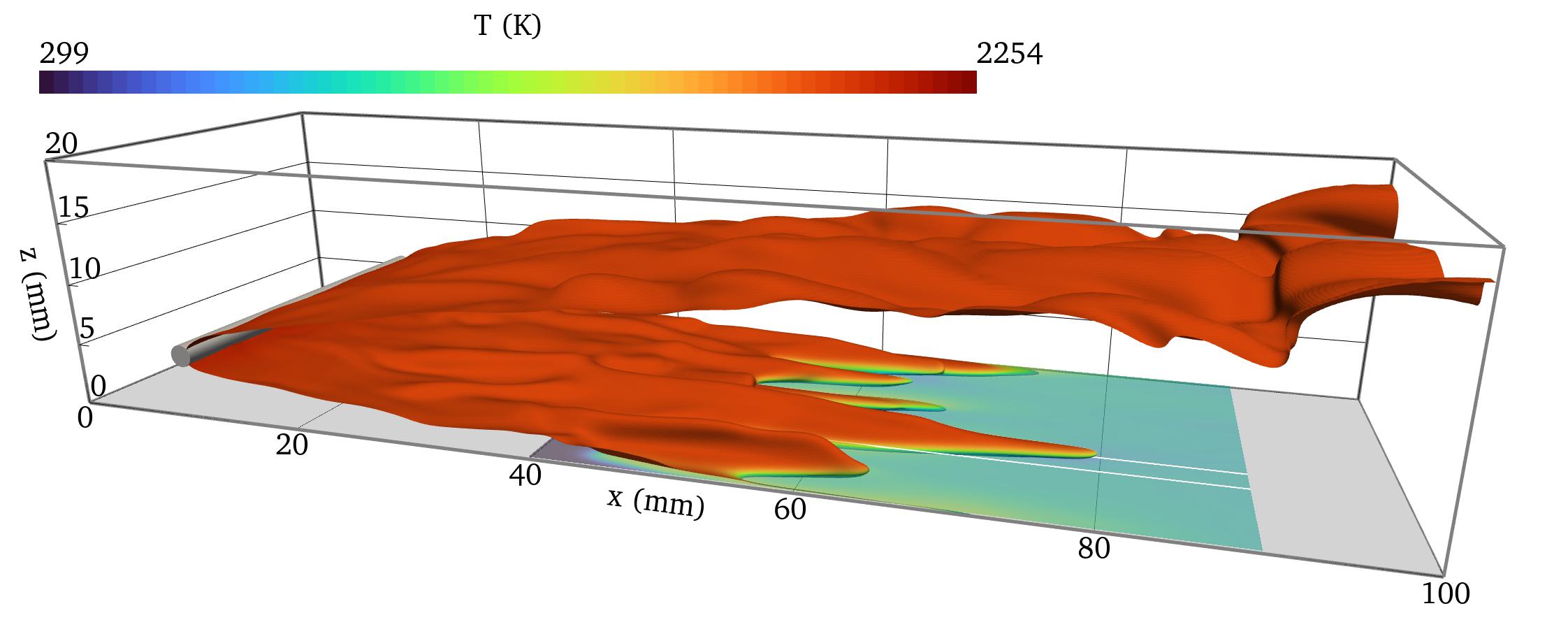}
  \caption{Snapshot of the reactive flow simulation. The flame surface is visualized by the iso-contour of $Y_{\text{CO}_\text{2}}=0.1$ and colored by temperature. The flame holder is indicated by the gray cylinder. At $z=0~\mathrm{mm}$ the wall is shown in gray. The simulation domain is shown as a gray bounding box. The slice close to the bottom wall corresponds to the area depicted in Fig.~\ref{fig:LES-sampling} and is colored by temperature.}
  \label{fig:snapshot}
\end{figure}

The quasi-DNS consists of two parts: a non-reactive simulation of the turbulent channel flow to generate appropriate turbulent inflow conditions and a reactive simulation of the V-shaped flame. The simulation domain of the non-reactive case is a channel with a length of $14H$, $3H$ and $2H$ in stream-wise ($x$), lateral ($y$) and wall-normal ($z$) direction, respectively. The flow is periodic in stream-wise and lateral direction, with a no-slip boundary condition applied at the walls. The computational mesh consists of 60 million hexahedral cells and is refined towards the walls, with a minimum grid size of $25~\text{\textmu m}$ or $y^+=0.24$. The inflow velocity fields at the boundary are stored at an interval corresponding to the simulation time step of $3~\text{\textmu s}$ and serve as an inflow condition for the reactive case. In the reactive simulation, the channel dimensions match the non-reactive counterpart, except for the channel length, which is reduced to $10H$. The computational mesh consists of 200 million cells (purely hexahedral, orthogonal mesh) refined towards the bottom wall with a minimum grid size of $12~\text{\textmu m}$ at the wall in the wall-normal direction or $y^+=0.14$. In this region, the Kolmogorov length scale $\eta$ has a minimum value of $45~\text{\textmu m}$ and the laminar flame thickness of the methane-air flame is
\begin{equation}
\delta_L = \frac{T_\text{burnt} - T_\text{unburnt}} {\max\left( \frac{\delta T}{\delta x} \right)} \approx 0.5~\text{mm} \ ,
\end{equation}
\noindent ensuring a sufficient grid resolution. The fine resolution near the bottom wall is not motivated by the resolution of the Kolmogorov length or the flame thickness, but by the sufficient resolution of the FWI zone, because the flame can move as close as $100~\text{\textmu m}$ toward the cold wall~\cite{zirwes2021numerical}. In the center of the domain (height of 1~cm), the wall-normal resolution is $100~\text{\textmu m}$ and the Kolmogorov length at that position is $150~\text{\textmu m}$. The velocity fields at the inlet boundary generated by the inert channel flow simulation are spatially and temporally interpolated to the inlet boundary face at every time step of the reactive simulation, which is about $\Delta t=0.3~\text{\textmu s}$ or $\text{CFL}=0.15$. In the lateral direction, periodic boundary conditions are applied. At the outlet a zero-gradient boundary condition is used for the velocity and the reactive scalars, while a Dirichlet boundary condition is employed for the pressure. The molecular diffusion coefficients for all species are assumed to be equal using a unity Lewis number assumption. The employed reaction mechanism is a reduced version of the CRECK mechanism~\cite{Ranzi2012} consisting of 24 species and 165 chemical reactions. The most important parameters of the numerical setup of the reactive simulation are summarized in Tab.~\ref{tab:setup}. Figure~\ref{fig:snapshot} shows the instantaneous flame front of the reactive simulation, depicted by the $Y_{\text{CO}_\text{2}}=0.1$ isoline.

The quasi-DNS are performed with an in-house solver~\cite{Zirwes2018,zirwes2018improved} implemented in the open-source CFD code OpenFOAM~\cite{Weller2017}, Version v1712. The solver uses finite-rate chemistry and was validated to be suitable for quasi-DNS in~\cite{Zirwes2018} using multiple DNS reference cases from the literature. The spatial discretization is based on a fourth-order interpolation scheme, and a second-order fully implicit backward scheme is used for the temporal discretization. The solver was used successfully for quasi-DNS of turbulent flames, examples are reported in~\cite{Zirwes2020,Hansinger2020,Zirwes2021a}. The reactive quasi-DNS was averaged over the duration of 20 flow trough times $20 \cdot (10H)/U_\mathrm{bulk} \approx 20 \cdot 0.02~\text{s}$. The simulations were performed on approximately 32,000 cores and more than 18 million core-h have been consumed.

\begin{table}[htb]
  \centering
  \caption{Numerical setup of the reactive case}
  \label{tab:setup}
  \small
  \begin{tabular}{ll}
      \hline
      \textbf{Parameter} & \textbf{Property} \\
      \hline
      Gas mixture & Stoichiometric methane-air flame \\
      Reaction mechanism & Reduced CRECK~\cite{Ranzi2012} \\
      Species diffusion model & Unity Lewis number transport \\
      Dimensions ($x \times y \times z$) & $100~\mathrm{mm} \times 30~\mathrm{mm} \times 20~\mathrm{mm}$ \\
      Flame anchor position (axial, wall-normal) & ($10~\mathrm{mm}$, $5~\mathrm{mm}$) \\
      Flame anchor radius & $0.9~\mathrm{mm}$ \\
      Mean inflow velocity & $4.4~\mathrm{m/s}$ \\
      Gas inlet temperature & $300~\mathrm{K}$ \\
      Wall temperature & $300~\mathrm{K}$ \\
      Reynolds number & $2770$ \\
      \hline
  \end{tabular}
\end{table}

\section{Modeling of turbulent flames using chemistry manifolds}

In the context of turbulent combustion modeling  using chemistry manifolds two major challenges need to be tackled; (i) a suitable chemistry manifold needs to be found that is able to reproduce the flame structure, and (ii) a TCI closure model is necessary that accounts for the unresolved contributions of the reactive scalars which in this particular context is the closure of the joint PDF. While the generation of a suitable manifold for FWI has been addressed in multiple studies of laminar~\cite{Ganter2017, Ganter2018, Strassacker2019a, Efimov2019b} and turbulent~\cite{Heinrich2018a, Heinrich2018b, Wu2015} flames, the TCI closure in the close vicinity of the wall has received less attention. This work focuses primarily on the closure of the joint PDF, hence, a state-of-the-art manifold is used. Its suitability is demonstrated in~\ref{sec:QFM-validation}.

The remainder of this section first introduces the manifold used in this work. Then, the modeling approaches for unresolved contributions of the reactive scalars are discussed in the context of both RANS and LES. Finally, the coupled {CQMOM} closure model employed in this work is presented.

\subsection{Chemistry manifolds for FWI}
\label{sec:QFM}
Chemistry manifolds for FWI have been validated in multiple studies of laminar side-wall quenching flames~\cite{Ganter2018, Steinhausen2021, Strassacker2019a} using detailed chemistry simulations and experiments. However, a comparison of the tabulated thermochemical states of the manifolds with a (quasi-)DNS of a turbulent side-wall quenching flame has not been investigated, yet. In this work, a two-dimensional Quenching Flamelet-Generated Manifold (QFM)~\cite{Efimov2019b} is used, a state-of-the-art manifold for FWI. It consists of a single, transient head-on quenching simulation and a series of preheated, one-dimensional freely propagating flames that extend the manifold at the upper enthalpy range. The simulation results are mapped on the normalized progress variable-enthalpy space $\phi = \phi \left( C, H \right)$, with the normalized enthalpy $H$ and progress variable $C$ given by
\begin{align}
  H &= \frac{h - h_\mathrm{min} }{ h_\mathrm{max} - h_\mathrm{min} } \ , \\
  \label{eq:C}
  C &= \frac{ Y_\mathrm{c} - Y_\mathrm{c, min} \left( H \right) }{ Y_\mathrm{c, max} \left( H \right) - Y_\mathrm{c, min} \left( H \right) } \ .
\end{align}
\noindent In the equation above, $h_\mathrm{min}$ and $h_\mathrm{max}$ are the minimum and maximum enthalpy in the chemistry manifold, respectively, while $Y_\mathrm{c, min} \left( H \right)$ and $Y_\mathrm{c, max} \left( H \right)$ are the minimum and maximum progress variable for a given enthalpy level. The progress variable $Y_\mathrm{c}$ is defined as the mass fraction of $Y_{\text{CO}_\text{2}}$. Further details regarding the thermochemical manifold and the table generation can be found in~\cite{Steinhausen2021, Efimov2019b}. The suitability of the manifold for turbulent side-wall quenching flames is studied in~\ref{sec:QFM-validation}.

\subsection{Closure of the joint PDF in the context of RANS and LES}
In the context of RANS/LES averaged/filtered balance equations are solved. Thereby, the local fluctuations and turbulent structures are integrated into mean/filtered quantities and do not need to be resolved in the simulation~\cite{Vervisch2002}. In the simulation of turbulent reacting flows instead of the commonly used Reynolds-average usually a Favre-average (mass-weighted) is used. In this work, the notation from~\cite{Vervisch2002} is employed, with $\overline{Q}$ and $Q'$ denoting the Reynolds mean and fluctuations of a quantity $Q$, respectively, while $\widetilde{Q}$ and $Q''$ correspond to the Favre-averaged counterparts. In the context of RANS, the averaging corresponds to a temporal average. Note that in the investigated channel flow, the temporal averaging is additionally performed in the statistical independent lateral direction $y$. In the context of LES, a spatial filter operation is performed using a box filter. A detailed description of the operations performed in the context of RANS and LES can be found in~\ref{appendix:averaging}.

To capture the correct behavior of the turbulent flame, the unresolved fluctuations of the reactive scalars can be described by their joint probability density function (PDF) that accounts for the temporal statistic in the flow in the context of RANS and the filtered density function (FDF) that considers the spatial fluctuations in the context of LES.
For the two-dimensional QFM employed in this work the bivariate PDF/FDF of the progress variable and enthalpy needs to be accounted for. In the following, both PDF and FDF are denoted by $\widetilde{P}$. Given the PDF/FDF $\widetilde{P} \left( Y_\mathrm{c}, H \right)$ of the progress variable $Y_\mathrm{c}$ and normalized enthalpy $H$ a mean/filtered quantity $\widetilde{Q}$ can be calculated in the context of RANS/LES

\begin{equation}
  \widetilde{Q} = \int_{Y_\mathrm{c}} \int_{H} f_Q \left( Y_\mathrm{c}, H \right) \widetilde{P} \left( Y_\mathrm{c}, H \right) dY_\mathrm{c}dH \ .
\end{equation}

\noindent Here $f_Q$ is a function that describes the dependency of $Q$ on the progress variable and normalized enthalpy (i.e. the thermochemical manifold) and $\widetilde{P}$ denotes the Favre-weighted PDF/FDF, which can be determined from the non-density weighted PDF/FDF $P$ as

\begin{equation}
  \widetilde{P} \left( Y_\mathrm{c}, H \right) = \frac{\rho}{\overline{\rho}} P \left( Y_\mathrm{c}, H \right) \ .
\end{equation}

\subsection{Closure of the joint PDF with CQMOM}
\label{sec:QMOM}

The Quadrature Method of Moments (QMOM)~\cite{Mcgraw1997} is based on Gaussian quadrature and approximates the unclosed integrals in the moment transport equations containing the unknown PDF/FDF. The approximation is accurate up to the order $2N-1$ with $N$ being the number of integration nodes employed. In this work a Conditional Quadrature Method of Moments (CQMOM)~\cite{Cheng2010} is employed to model the moments of the bivariate PDF/FDF $\widetilde{P} \left( Y_\mathrm{c}, H \right)$; this represents an extension of the standard (univariate) QMOM approach~\cite{Mcgraw1997} to multivariate PDFs/FDFs based on the concept of a conditional PDF/FDF~\cite{Yuan2011}, i.e. the bivariate PDF/FDF is given as

\begin{equation}
  \label{eq:CQMOM_PDF}
  \widetilde{P} \left( Y_\mathrm{c}, H \right) = \widetilde{P} \left( Y_\mathrm{c} \right) P \left( H | Y_\mathrm{c} \right) \ ,
\end{equation}

\noindent where $\widetilde{P} \left( Y_\mathrm{c} \right)$ is the marginal PDF/FDF of the progress variable and $P \left( H | Y_\mathrm{c} \right)$ indicates the conditional PDF/FDF of the normalized enthalpy for a given value of the progress variable. The CQMOM uses a set of primary $\phi_{\alpha}$ and secondary (conditional) $\phi_{\alpha;\beta}$ nodes to approximate the joint PDF. The subscript $\alpha$ hereby describes the index of the primary direction, while $\left( \alpha;\beta \right)$ reads as index $\beta$ of the secondary direction for a given index in the primary direction $\alpha$. In the approximation, for each $N_\alpha$ nodes in the primary direction, $N_\beta$ nodes in the secondary direction are defined that model the conditional PDF/FDF $P \left( H|Y_\mathrm{c} \right)$. Thus, a single node in secondary direction implies, that for every integration node in primary direction a single integration node in secondary direction is used. In this work, the primary and secondary (conditional) direction correspond to the progress variable $Y_\mathrm{c}$ and normalized enthalpy $H$, respectively. The integral approximation can be written as

\begin{equation}
\int_{\mathcal{R}_\psi} q \left( \phi_{\alpha}, \phi_{\alpha;\beta} \right) \widetilde{P} \left( \phi_{\alpha}, \phi_{\alpha;\beta} \right) d\phi_{\alpha} d\phi_{\alpha;\beta} \approx \sum_{\alpha=1}^{N_\alpha} \sum_{\beta=1}^{N_\beta} w_{\alpha}w_{\alpha;\beta} q\left( \phi_{\alpha}, \phi_{\alpha;\beta}  \right) \ ,
\end{equation}

\noindent with $w_{\alpha}$ and $w_{\alpha;\beta}$ being the primary and conditional weights of the bivariate PDF/FDF and $q\left( \phi_{\alpha}, \phi_{\alpha;\beta} \right)$ containing all terms except the PDF/FDF itself. Using $q\left( \phi_{\alpha}, \phi_{\alpha;\beta} \right) = \phi_{\alpha}^k \phi_{\alpha;\beta}^l $ yields the moment definition

\begin{equation}
\widetilde{m}_{k, l} = \int_{\mathcal{R}_\psi} \phi_{\alpha}^k \phi_{\alpha;\beta}^l \widetilde{P} \left( \phi_{\alpha}, \phi_{\alpha;\beta} \right) d\phi_{\alpha} d\phi_{\alpha;\beta} \approx \sum_{\alpha=1}^{N_\alpha}\sum_{\beta=1}^{N_\beta} w_{\alpha}w_{\alpha;\beta} \phi_{\alpha}^k \phi_{\alpha;\beta}^l \ .
\end{equation}

\noindent To perform the moment conversion (calculation of integration nodes and weights) a system of moments is used that are solved for in a coupled CFD simulation.
To reach a fully determined system of moments for the bivariate PDF/FDF, a minimum of $2 \cdot N_{\alpha}$ primary moments and $N_{\alpha} \cdot (2 \cdot N_{\beta} - 1)$ conditional moments are necessary. For additional information on the system of moments, the reader is referred to~\cite{Marchisio2013}. The nodes and weights of the PDF/FDF are computed using a two-stage Wheeler algorithm that is described in~\cite{Wheeler1974, Marchisio2013}. First, the primary nodes $\phi_\alpha$ are computed corresponding to the progress variable direction. The results are then used to calculate the nodes of the conditional moments in the enthalpy direction $\phi_{\alpha;\beta}$. Finally, the PDF/FDF is represented as a weighted sum of Dirac delta functions

\begin{equation}
    \widetilde{P} \left( Y_\mathrm{c}, H \right) \approx \sum_{\alpha=1}^{N_\alpha} \sum_{\beta=1}^{N_\beta} w_{\alpha} w_{\alpha;\beta} \delta \left(Y_\mathrm{c} - \phi_{\alpha} \right) \delta \left(H - \phi_{\alpha; \beta}\right) \ .
\end{equation}

\noindent Note that in the equation above, the PDF/FDF is modeled using a conditional PDF/FDF in the secondary direction, since the nodes in the secondary (conditional) direction are defined for every primary node separately, indicated by the subscript $\left( \alpha; \beta \right)$. An integral quantity $\widetilde{Q}$, can then be approximated by

\begin{equation}
  \label{eq:mean_quantity}
  \widetilde{Q} \approx \sum_{\alpha=1}^{N_\alpha}\sum_{\beta=1}^{N_\beta} w_{\alpha}w_{\alpha;\beta} f_Q \left( \phi_{\alpha}, \phi_{\alpha;\beta} \right) \ .
\end{equation}

\noindent In a coupled simulation, the algorithm presented in~\cite{MadadiKandjani2017, Fox2018, Pollack2021} can be used to solve the moment transport equations.

\section{PDFs and FDFs in the context of flame-wall interaction}
\label{sec:PDFs}

In the following, the particular challenges of the closure of the joint PDF in the context of FWI are examined. First, the joint PDF of the progress variable and enthalpy to be modeled in the context of RANS is examined for various wall distances. Secondly, selected FDFs at specific flame positions are discussed in the context of LES.

\subsection{Analysis of the joint PDF in the context of RANS}
\label{sec:PDFs-RANS}

To calculate the joint PDF the quasi-DNS is sampled with a time-step of $\Delta t = 2.5 \cdot 10^{-4}~\text{s}$ over the duration of two flow-through times ($0.04~\text{s}$). Additionally, the data is averaged over the statistically independent lateral channel dimension. The PDF is extracted at different positions $\left( x_i, z_i \right)$ defined by the wall-normal distance $z$ and a stream-wise location $x$:

\begin{itemize}
  \item \textbf{Wall-normal distance ($\mathbf{z}$):} Different wall-normal distances are examined to investigate the impact of wall-heat losses on the joint PDF.

  \item \textbf{Stream-wise location ($\mathbf{x}$):} Different stream-wise locations are extracted to analyze the influence of the reaction progress on the PDF. The stream-wise location is defined for every wall-normal position, separately, and is based on the time-averaged normalized progress variable $\overline{C}$.
\end{itemize}

\begin{figure}[htb]
  \centering
  \includegraphics[scale=1.0]{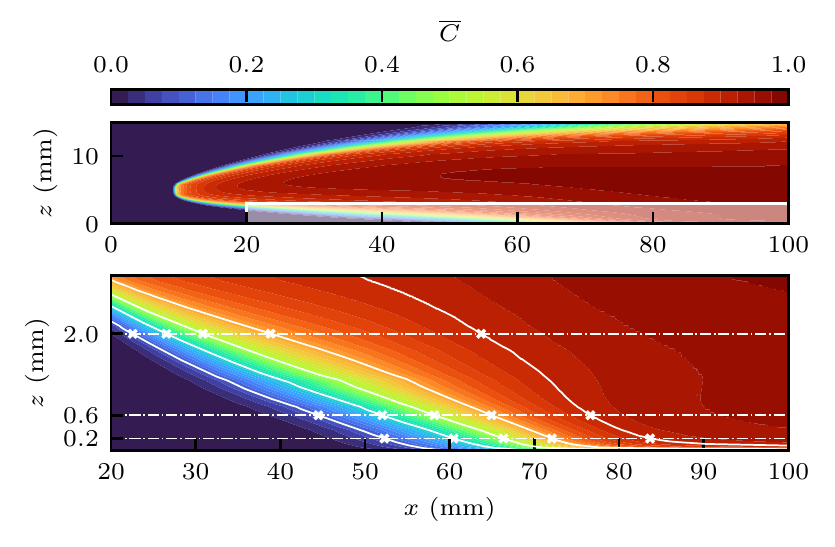}
  \caption{Contour plot of the time-averaged normalized progress variable $\overline{C}$ (top) and a zoomed image of the near-wall region (bottom). The white rectangle in the top plot shows the magnified region in the bottom. In the zoomed image, the iso-contours of $\overline{C} = \left[ 0.1, 0.3, 0.5, 0.7, 0.9 \right]$ are shown as solid lines. The dash-dotted lines correspond to $z=[0.2, 0.6, 2]~\mathrm{mm}$. The markers indicate the extraction points for the PDFs that are shown in Fig.~\ref{fig:RANS-pdfs}.}
  \label{fig:RANS-sampling}
\end{figure}

Figure~\ref{fig:RANS-sampling} shows a contour plot of the time-averaged normalized progress variable $\overline{C}$ (top) and a zoomed image of the near-wall region (bottom). In the zoomed image the sample positions for data extraction are indicated by the markers and the wall-normal extraction heights are shown as dash-dotted lines. Note that for the methane-air flame ($\delta_L\approx0.5~\text{mm}$) investigated in this work the wall distance normalized by the laminar flame thickness $z/\delta_L$ is equivalent to approximately twice the wall distance $z$.

\newpage

\begin{figure}[htb]
  \includegraphics[scale=1.0]{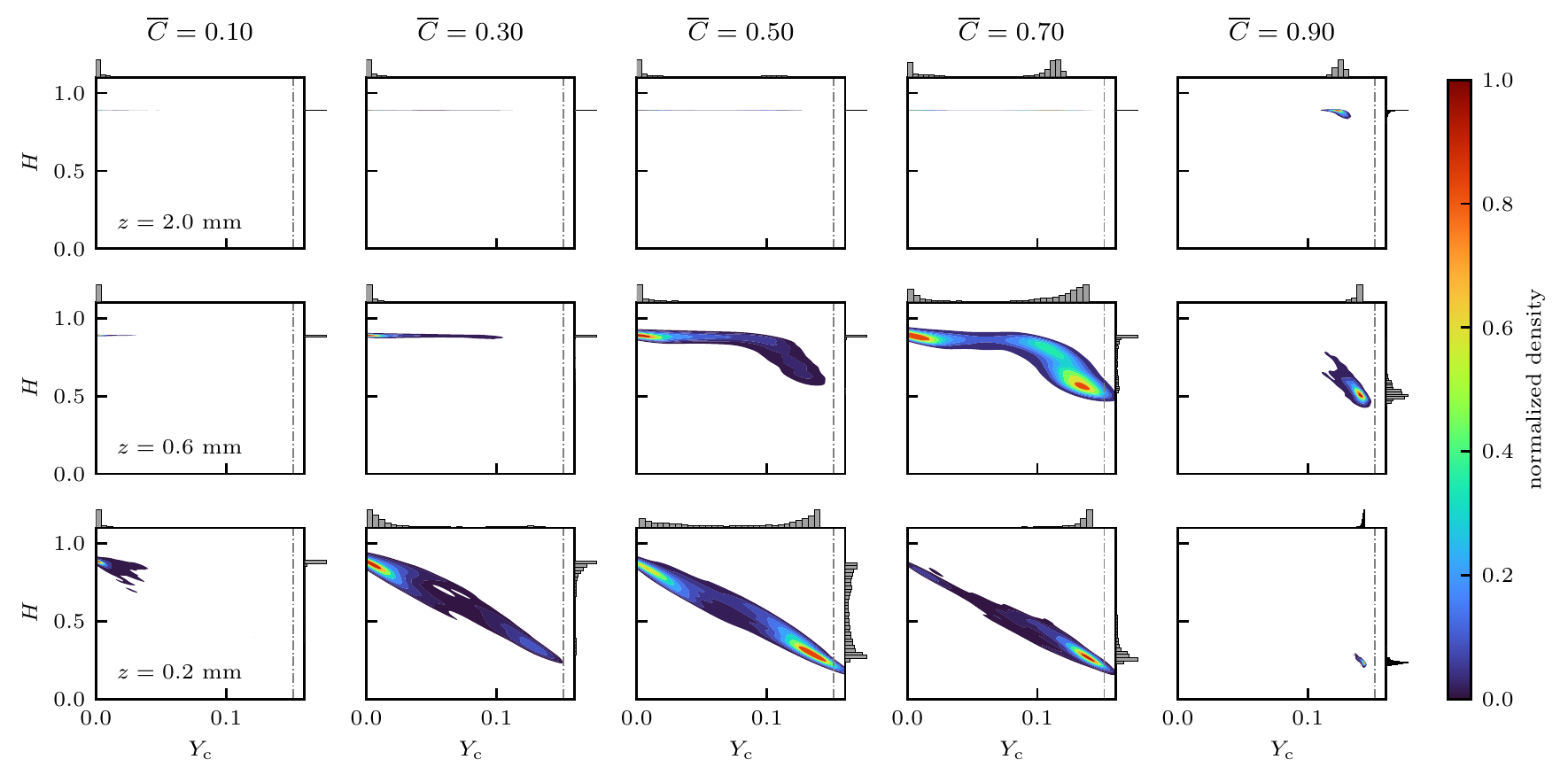}
  \caption{PDFs at a wall distance of $z=2.0~\text{mm}$ (top) $z=0.6~\mathrm{mm}$ (middle) and $z=0.2~\text{mm}$ (bottom). Each subplot shows the joint PDF of the progress variable  $Y_\mathrm{c}$ and normalized enthalpy $H$ as a contour colored according to the probability normalized by its maximum. The marginal PDF of the progress variable (top) and normalized enthalpy (right) are shown at the figure borders as bar plots with a bin size of 20. The mean reaction progress increases from left to right. The dash-dotted line corresponds to the global maximum value of $y_c$ and the minimum is 0.}
  \label{fig:RANS-pdfs}
\end{figure}

Figure~\ref{fig:RANS-pdfs} shows the extracted temporal joint PDFs of the progress variable and normalized enthalpy and their respective marginal PDFs at the figure borders for the progress variable (top) and normalized enthalpy (right). First, the joint PDFs are discussed with respect to the wall distance~$z$. Therefore, the isoline $\overline{C}=0.5$ is considered and the plots are analyzed from the core flow (top) to the close vicinity of the wall (bottom). In the core flow ($z=2.0$~mm, top row), the flame is unaffected by enthalpy losses to the wall and a univariate PDF solely dependent on the progress variable can be observed, i.e. the enthalpy is constant as expected for a unity Lewis number case. With decreasing wall distance~$z$, the PDF shape changes. Due to enthalpy losses to the wall the PDF becomes bivariate varying with both progress variable and enthalpy. At $z=0.6~\mathrm{mm}$ (middle row), only high values of the progress variable ($Y_\mathrm{c}>0.1$) are significantly affected by the wall heat losses, showing a decrease in normalized enthalpy. In these regions, the PDF broadens in enthalpy direction. For lower progress variables ($Y_\mathrm{c}<0.1$), however, the PDF retains its univariate character of the core flow. Finally, in the close vicinity to the wall ($z=0.2~\mathrm{mm}$, bottom row), a bivariate PDF can be observed for all values of the mean progress variable.

Secondly, the influence of mean reaction progress on the PDF is discussed at different stream-wise locations shown in Fig.~\ref{fig:RANS-pdfs} from left to right. At the unburnt ($\overline{C}=0.1$) edges of the flame the fluctuations in both the progress variable and enthalpy direction vanish and the PDF can be approximated by a single point in the progress variable-enthalpy space. With increasing reaction progress, first the PDF widens in progress variable direction, still showing mainly low progress variable values ($\overline{C}=0.3$). Then, a typical double peak PDF of the progress variable can be observed ($\overline{C}=0.5/0.7$), showing a high probability of low and high values of the progress variable and few occurrences of intermediate values. This shape of the distribution is a direct effect of the thin reaction zones in premixed flames. With further increasing reaction progress, the PDF shows mainly high values of progress variable, before it approaches a single peak PDF with a small variance towards the burnt state.

\begin{figure}[htb]
  \centering
  \includegraphics[scale=1.0]{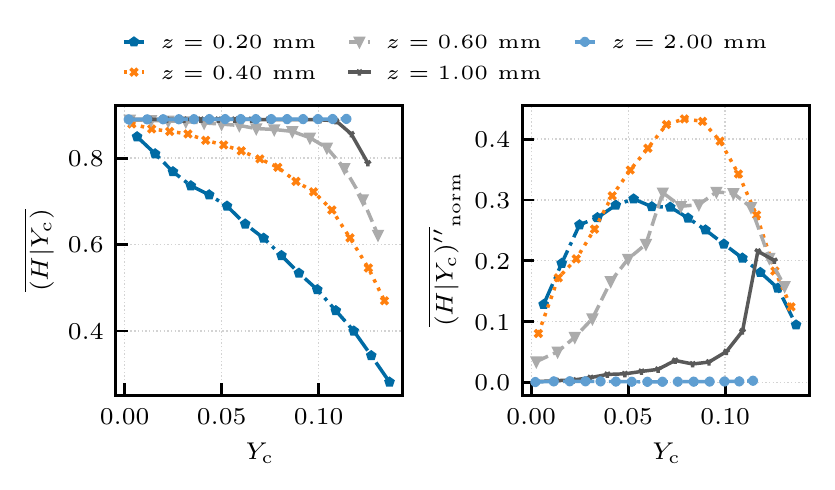}
  \caption{Conditional mean and normalized standard deviation of the normalized enthalpy for a given progress variable at different wall distances $z$ and $\overline{C}=0.5$.}
  \label{fig:RANS-cond-pdf}
\end{figure}

In summary, the influence of enthalpy losses at the wall increases with both, increasing reaction progress and decreasing wall distance due to higher temperature gradients in the flow. These enthalpy losses in the near-wall region yield a complex bivariate PDF in the reaction zone of the flame that need to be accounted for in TCI closure. The trends of the joint PDFs are also reflected in their marginal counterparts leading to a broadening of the univariate distributions of the progress variable and enthalpy. However, the correlation of the progress variable and enthalpy, as e.g. observed for $z=0.2$ mm, is not captured by the marginal PDFs. In particular, statistical independence~\cite{Fiorina2005, Donini2017, Zhang2021} as often used in presumed PDF methods, is not applicable for FWI. In this context, the CQMOM approach that models the joint PDF as a marginal PDF of the progress variable and a conditional PDF of the enthalpy, see Eq.~\eqref{eq:CQMOM_PDF} is advantageous. Figure~\ref{fig:RANS-cond-pdf} shows the conditional mean and normalized standard deviation of the normalized enthalpy for a given progress variable at different wall distances $z$. The conditional quantities clearly reflect the correlation of progress variable and enthalpy in the joint PDFs, indicating the benefit of the CQMOM method over the pPDF approach. Additionally, the wall normal distance affected by enthalpy losses to the wall can be deduced from the left plot in Fig.~\ref{fig:RANS-cond-pdf}. While for laminar flames~\cite{Zhang2021} an influence of the wall on the flame can be observed for $z/\delta_L < 1$, in the turbulent flame investigated here $z/\delta_L < 2$.

\subsection{Analysis of the joint FDF in the context of LES}
\label{sec:PDFs-LES}

\begin{figure}[htb]
  \centering
  \includegraphics[scale=1.0]{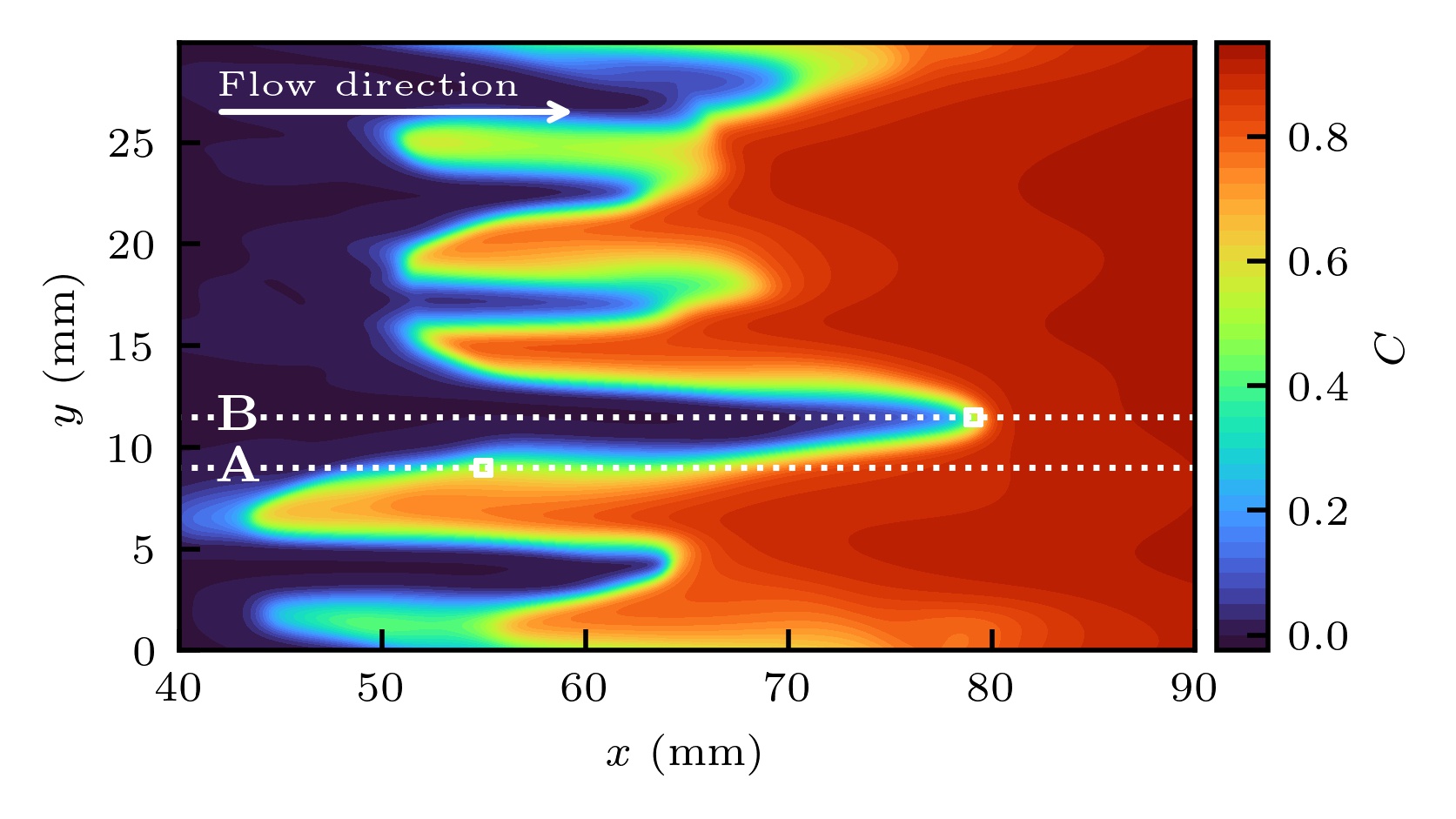}
  \caption{Snapshot of a wall-parallel cut through the domain at $z=0.2~\text{mm}$ colored by the normalized progress variable $C$. Two representative flame positions are indicated by the dotted lines: flame flank (A), flame tip (B). The markers depict the center points of the box-samples used to calculate the FDFs. The slice is also shown in the 3D view of the flame in Fig.~\ref{fig:snapshot}.}
  \label{fig:LES-sampling}
\end{figure}

In the context of LES, the flame is filtered locally for instantaneous fields using a box filter operation (Eq.~\eqref{eq:box-filter}). Therefore, transient processes in the flame affect the FDF and need to be considered in the analysis. In Fig.~\ref{fig:snapshot} a snapshot of the DNS is shown and distinctive flame tongues can be observed at different lateral positions. Over time, these flame tongues emerge at different lateral locations in the flame, slowly propagate forward and are finally rapidly pushed back to the position of the main reaction front. This results in a locally high-intensity wall heat flux. A similar behavior is also described in~\cite{Gruber2010} for a hydrogen-air flame and in~\cite{Heinrich2018a} for a methane-air flame and can be explained by the interaction of the flame with the near-wall vortices. Figure~\ref{fig:LES-sampling} shows a wall-parallel cut at $z=0.2~\mathrm{mm}$ of the flame snapshot in Fig.~\ref{fig:snapshot}. In the figure, two representative flame zones are depicted as dotted lines: a flame flank (A) and a flame tip (B). The flame flank (A) is characterized by a position in the flame with a high progress variable gradient in the lateral direction, while the flame tip (B) corresponds to the maximum of the flame front in the stream-wise direction. Furthermore, similar to Fig.~\ref{fig:RANS-sampling}, the box filter's center points of the FDFs analyzed in the following are shown. The center points ($x_{i,\text{center}}$, $y_{i,\text{center}}$, $z_{i,\text{center}}$) are determined as follows:

\begin{itemize}
    \item \textbf{Wall-normal distance ($\mathbf{z}_\text{center}$):} Different wall-normal distances are examined to investigate the impact of wall-heat losses on the joint FDF.
    \item \textbf{Lateral location ($\mathbf{y}_\text{center}$):} Other than in the context of RANS, in LES the lateral direction is relevant for the flame, since the data is not averaged over time, but resolved locally (3D). Two respective lateral positions are chosen corresponding to a flame tip and a flame flank.
    \item \textbf{Stream-wise location ($\mathbf{x}_\text{center}$):} Different stream-wise locations are extracted to analyze the influence of the reaction progress on the FDF. The stream-wise location is defined for every wall-normal and lateral position, separately, and is based on the instantaneous normalized progress variable $C$. Different stream-wise locations were analyzed that showed the highest challenges for the closure of the joint PDF in the reaction zone of the flame. In the following, only a representative position is discussed with $C=0.5$. At the other stream-wise locations similar observation can be made.
\end{itemize}

In the following analysis different exemplary filter kernels are employed to show the influence of the LES resolution on the FDF closure. The filter kernel size is given by

\begin{equation}
    \label{eq:box-filter-size}
    \left( \Delta_x, \Delta_y, \Delta_z \right) = \left( \mathrm{AR} \cdot\delta_z, \mathrm{AR} \cdot\delta_z, \delta_z \right) \ ,
\end{equation}

\noindent with $\delta_z$ being the filter width in wall-normal direction and $\mathrm{AR}$ the aspect ratio of the filter kernel, indicating the influence of a grid refinement at the wall in an LES simulation. Note that even though the mesh is (formally) unstructured, due to its purely hexahedral and orthogonal structure the application of a box filter is a straight forward operation. In the following, first the FDF is discussed for different lateral positions and wall normal filter width $\delta_z$. Then, the influence of the apect ratio is analyzed.

\begin{figure}[htb]
  \includegraphics[scale=1.0]{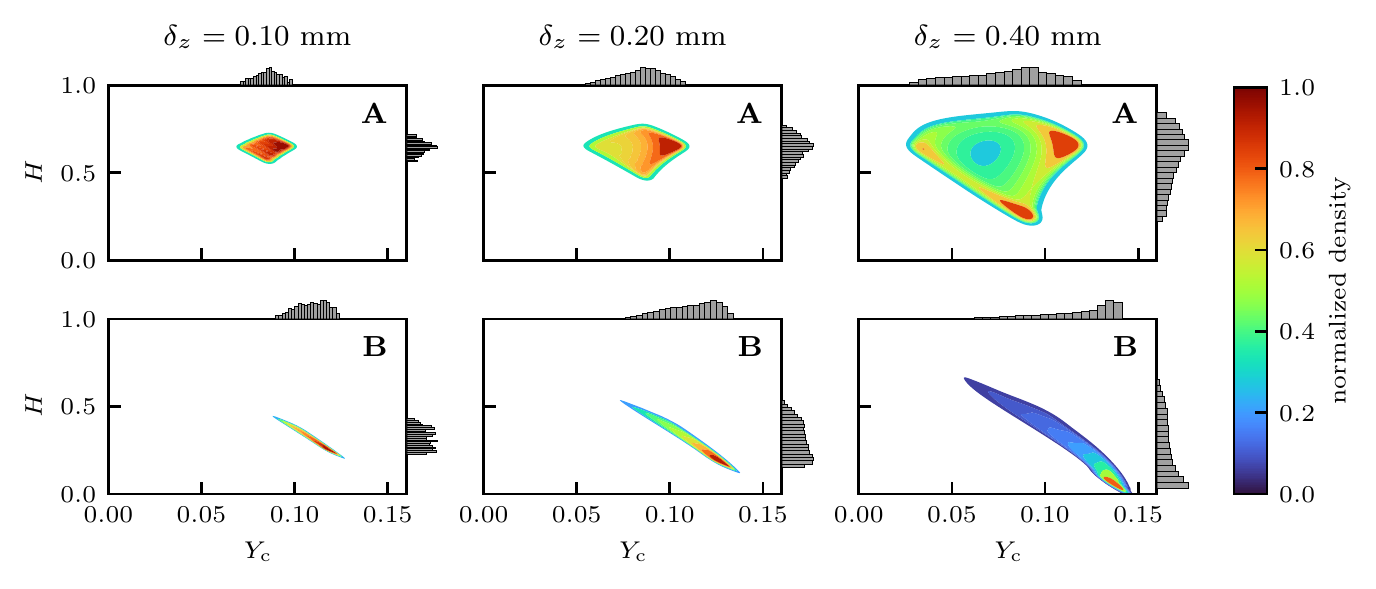}
  \caption{FDFs in the context of LES at a wall distance of $z=0.2~\text{mm}$ and a normalized progress variable of $C=0.5$. At the top the FDFs at the flame flank (A) are shown, while at the bottom the FDFs at the flame tip (B) are depicted. From left to right the box-size (LES filter width) increases. The aspect ratio of the box filter is $\mathrm{AR}=2.5$. Each subplot shows the joint FDF of the progress variable  $Y_\mathrm{c}$ and normalized enthalpy $H$ as a contour (middle) colored according to the probability normalized by its maximum. The marginal FDF of the progress variable (top) and normalized enthalpy (right) are shown at the figure borders as bar plots with a bin size of 20.}
  \label{fig:LES-pdfs}
\end{figure}

The marginal and joint Favre-weighted FDFs of the progress variable and enthalpy are shown in Fig.~\ref{fig:LES-pdfs} for different $\delta_z$ and $\mathrm{AR}=2.5$. The flame flank (A) shows a wide distribution for the progress variable and enthalpy and no significant correlation between the two variables can be observed. The flame tip (B), on the other hand, shows a line-like distribution that indicates a strong correlation between the progress variable and enthalpy in this area of the flame. These two completely different FDF shapes show the challenge for any TCI closure model for FWI, which must be able to describe both FDF shapes. Note that the marginal FDFs (subfigure borders) that are used in pPDF approaches, do not reflect the correlations of progress variable and enthalpy in the joint FDFs. In the context of CQMOM, the FDFs at the flame tip (B) are straightforward due to the use of a conditional FDF for the enthalpy. For classical pPDF approaches from literature~\cite{Fiorina2005, Donini2017,Zhang2021}, however, the high correlation between progress variable and enthalpy is in contradiction to the modeling assumptions of independence of the progress variable and enthalpy. The flame flank (A), on the other hand, features a more challenging FDF for the CQMOM approach due to the low correlation between the progress variable and enthalpy and this is investigated in the following section.

Secondly, the impact of wall-normal filter width $\delta_z$ on the FDF complexity is discussed. In Fig.~\ref{fig:LES-pdfs} the filter width is increased from left to right by a factor of two for each subplot. While the FDF shape is mostly dependent on the specific flame area (flame flank/flame tip), the FDF widens with increasing filter width and shows increasing variance in the progress variable and enthalpy direction leading to a more complex FDF that needs to be accounted for by the TCI closure model. This aspect of TCI closure is also discussed in subsection~\ref{sec:CQMOM-TCI-LES}.

\begin{figure}[htb]
  \includegraphics[scale=1.0]{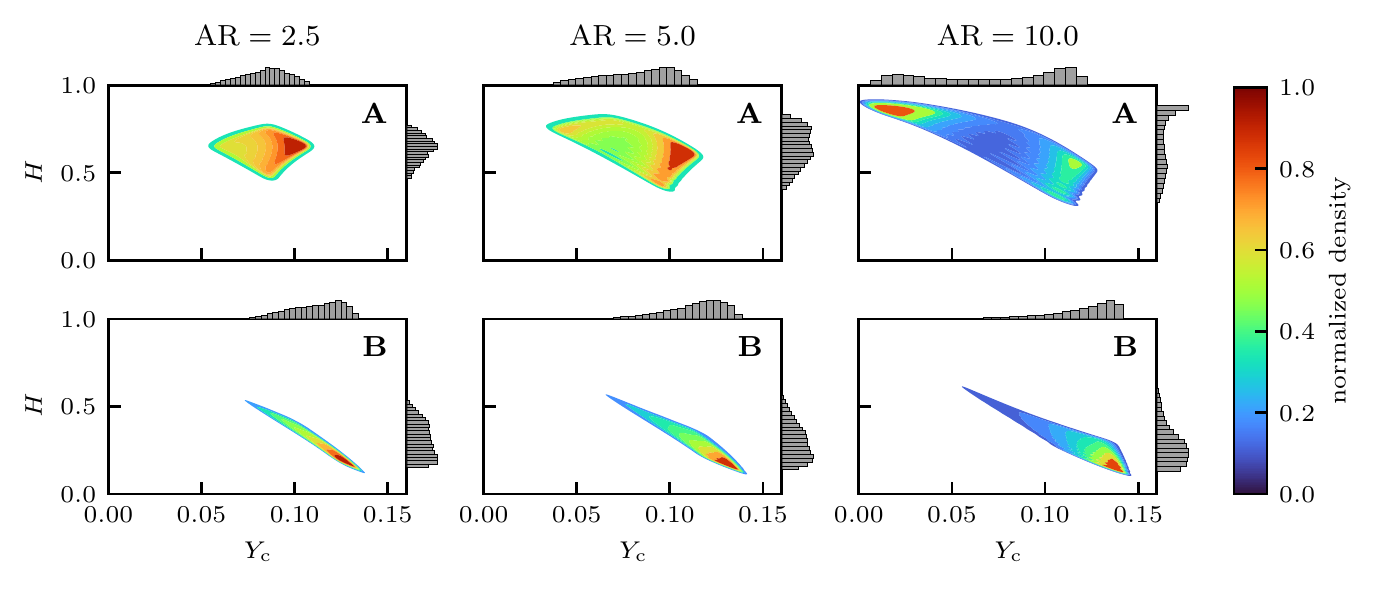}
  \caption{FDFs in the context of LES at a wall distance of $z=0.2~\text{mm}$ and a normalized progress variable of $C=0.5$. At the top the FDFs at the flame flank (A) are shown, while at the bottom the FDFs at the flame tip (B) are depicted. From left to right the aspect ratio increases. The wall-normal filter width is $\delta_z=0.2~\mathrm{mm}$. Each subplot shows the joint FDF of the progress variable $Y_\mathrm{c}$ and normalized enthalpy $H$ as a contour (middle) colored according to the probability normalized by its maximum. The marginal FDF of the progress variable (top) and normalized enthalpy (right) are shown at the figure borders as bar plots with a bin size of 20.}
  \label{fig:LES-pdfs-AR}
\end{figure}

Finally, the influence of the aspect ratio $\mathrm{AR}$ is discussed. Figure~\ref{fig:LES-pdfs-AR} shows the FDFs corresponding to $\delta_z = 0.2~\mathrm{mm}$ and varying aspect ratio. At the flame flank (A) the FDF is significantly influenced by the aspect ratio. While for small aspect ratios, a relatively uniform distribution is present, with increasing aspect ratio the FDF approaches a double peak FDF, similar to the ones observed in the context of RANS with a high probability of fresh and burnt gas states. The flame tip (B) on the other hand is not significantly influenced by the aspect ratio remaining its general shape, showing only a broadening of the FDF with increasing aspect ratio.

\reviewtwo{Note that the filter widths discussed above have a maximum value of $2~\mathrm{mm}$ (four time the laminar flame thickness $\delta_L$) in the stream-wise and lateral direction and $0.4~\mathrm{mm}$ in the wall-normal direction. Dependent on the configuration considered the LES filter widths can be even larger in a coupled simulation.}

\section{\textit{A priori} validation of the CQMOM closure of the joint PDF}
\label{sec:apriori}
In this section, the suitability of the CQMOM approach to model the unresolved PDFs/FDFs is assessed in an \textit{a priori} analysis and compared to a pPDF approach from the literature~\cite{Fiorina2005} that uses a $\beta$-PDF for the progress variable and a $\delta$-peak for the enthalpy. The analysis focuses on the PDFs/FDFs in the reaction zone. Here, the modeling challenges for the closure of the joint PDF are the highest due the very complex PDF/FDF shapes as discussed in the previous section.

\begin{figure}[htb]
  \includegraphics[width=14cm]{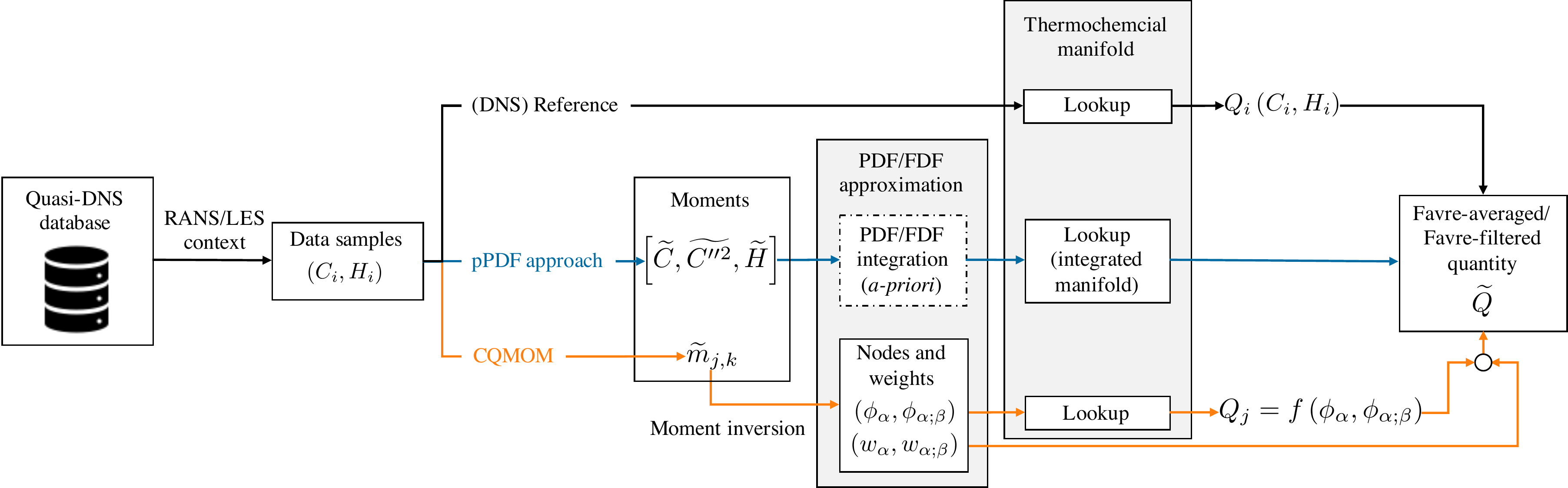}
  \caption{Schematic view of the \textit{a priori} analysis for a flow quantity $Q$. The steps of the analysis are illustrated from left to right, starting from the quasi-DNS database and resulting in an approximation of a Favre-averaged/Favre-filtered quantity $\widetilde{Q}$. In the context of RANS, a temporal average is performed, while in the context of LES, Favre-filtered quantities are considered. Note that the PDF/FDF approximation step of the pPDF approach can be performed prior to a simulation, while the CQMOM calculations of nodes and weights are performed during runtime.}
  \label{fig:apriori}
\end{figure}

It is important to note that this \textit{a priori} analysis differs from other analyses performed in the literature~\cite{Bray2006, Jin2008}, that directly compare the PDF/FDF shapes of the DNS reference with the presumed PDF/FDF counterpart. The CQMOM approach does not predict the PDF/FDF itself but instead estimates integrals based on the unknown PDF/FDF. Hence, the shape of the PDF/FDF cannot be directly analyzed (or compared) in the context of the QMOM method used here. Thus, the \textit{a priori} analysis examines the prediction accuracy of Favre-averaged/Favre-filtered quantities originating from the respective PDFs/FDFs. Such an approach is particularly advantageous for the latter use in LES or RANS. Figure~\ref{fig:apriori} illustrates the workflow of the \textit{a priori} analysis performed in this work.

To ensure a fully consistent comparison of the closure models of the joint PDF/FDF, all thermochemical quantities are taken directly from the QFM manifold, i.e. we also perform a lookup using the DNS values of progress variable and enthalpy. This effectively decouples the error in the manifold from the TCI analysis.
Starting from the quasi-DNS database, the data is sampled according to the respective context. In the context of RANS the Favre-averaged quantities are calculated on the same samples that are used to calculate a temporal PDF as described in Section~\ref{sec:PDFs-RANS}, while in the context of LES the filtered quantites are based on the FDF data extraction, described in Section~\ref{sec:PDFs-LES}. From these samples, the Favre-averaged/Favre-filtered quantities can be calculated as follows for the quasi-DNS, the pPDF approach and the CQMOM closure:

\begin{itemize}
  \item \textbf{(DNS) Reference:} The quasi-DNS reference values are calculated from the $N$ samples that estimate the unresolved PDFs/FDFs, see Section~\ref{sec:PDFs}:
  \begin{equation}
    \widetilde{Q}_\text{ref} = \frac{1}{\overline{\rho}} \sum_{i=0}^{N} \underbrace{f_\rho \left( C_{i}, H_{i} \right)}_\text{from QFM} \underbrace{f_Q \left( C_{i}, H_{i} \right) }_\text{from QFM} \ .
  \end{equation}
  Note that both the density and the quantity $Q$ are estimated from the thermochemical manifold indicated by $f_\rho$ and $f_Q$, respectively.
  \item \textbf{pPDF approach:} The approach by Fiorina et al.~\cite{Fiorina2005} using a $\beta$-PDF for the normalized progress variable and a $\delta$-peak for the normalized enthalpy results in:
  \begin{equation}
    \widetilde{Q} = f \left( \widetilde{C}, \widetilde{C''^{2}}, \widetilde{H} \right) = \int \int f_Q \left( C, H \right) \cdot \widetilde{P}\left( C \right) \widetilde{P} \left( H \right) dC dH \ ,
  \end{equation}
  \noindent where the $\beta$-PDF $\widetilde{P}\left( C \right)$ is defined by the first and second moment $\left( \widetilde{C}, \widetilde{C''^2} \right)$, $\widetilde{P} \left( H \right)\approx \delta \left( H - \widetilde{H} \right)$ is a $\delta$-PDF centered at $\widetilde{H}$.
  \item \textbf{CQMOM:} In the \textit{a priori} CQMOM closure, only part of the algorithm described in~\cite{MadadiKandjani2017, Fox2018, Pollack2021} needs to be used. In particular, instead of solving  transport equations for the moments, they are directly calculated from the quasi-DNS samples (step i. below) and used as input parameters for the CQMOM algorithm. The CQMOM approximation, for a Favre-averaged/Favre-filtered quantity $\widetilde{Q}$, is calculated in a multi-step process:
  \begin{enumerate}[i.]
    \item \textbf{Moment calculation:} The Favre-averaged/Favre-filted moments $\widetilde{m}_{k,j}$ are calculated directly from $N$ sample points acquired from the quasi-DNS result:
    \begin{equation}
      \widetilde{m}_{k,j}=\frac{ \sum_{i=0}^{N} f_\rho\left( C_{i}, H_i \right) Y_{\mathrm{c},i}^k H_i^j}{\sum_{i=0}^{N} \underbrace{f_\rho \left( C_{i}, H_i \right)}_{\text{from QFM}}} \ .
    \end{equation}
     \item \textbf{Moment inversion:} The nodes $\left( \phi_{\alpha}; \phi_{\alpha;\beta} \right)$ and weights $\left( w_{\alpha}; w_{\alpha;\beta} \right)$ corresponding to the given moment set $\widetilde{m}_{k,j}$ are calculated.
    \item \textbf{Table lookup:} For each node, the flow quantity can be extracted from the chemistry manifold $f_Q \left( \phi_{\alpha}, \phi_{\alpha;\beta} \right)$. Note here that the PDF/FDF nodes correspond to a point in the progress variable-enthalpy state.

    \item \textbf{Calculate the means:} Finally, the Favre-averaged/Favre-filtered quantity $\widetilde{Q}$ can be calculated using the interpolation weights, and nodes corresponding to the Favre-averaged moments using Equation~\eqref{eq:mean_quantity}.
  \end{enumerate}
\end{itemize}

For the analysis two flow quantities are examined that show a strong sensitivity to wall heat losses, namely the temperature and the mass fraction of CO. These two quantities have also been of central interest in multiple studies of laminar~\cite{Ganter2017, Ganter2018, Kosaka2018, Jainski2017, Jainski2017a} and turbulent~\cite{Jainski2018c, Kosaka2018} side-wall quenching. Even though the source term of the progress variable is a central quantity for the closure of the joint PDF it is not used for the analysis performed in the following, since (i) the suitability of the QMOM approach to predict the source term of the progress variable has been discussed in a previous study of laminar freely propagating flames~\cite{Pollack2021} and (ii) due to flame quenching the reaction in the flame stagnates in the close vicinity to the wall leading to a vanishing source term in the FWI area. In the following, after a convergence study in section~\ref{sec:CQMOM-convergence}, the CQMOM closure is discussed in the context of RANS (section \ref{sec:CQMOM-TCI-RANS}) and LES (section~\ref{sec:CQMOM-TCI-LES}).

\subsection{CQMOM moment convergence study}
\label{sec:CQMOM-convergence}
Before the CQMOM approach is compared to the pPDF approach from the literature the necessary number of integration nodes in the progress variable $n_{Y_\mathrm{c}}$ and normalized enthalpy $n_{H}$ direction needs to be assessed. Therefore, the prediction of the Favre-averaged/Favre-filtered CO mass fraction is analyzed in the context of RANS/LES.

\begin{figure}[htb]
  \includegraphics[scale=1.0]{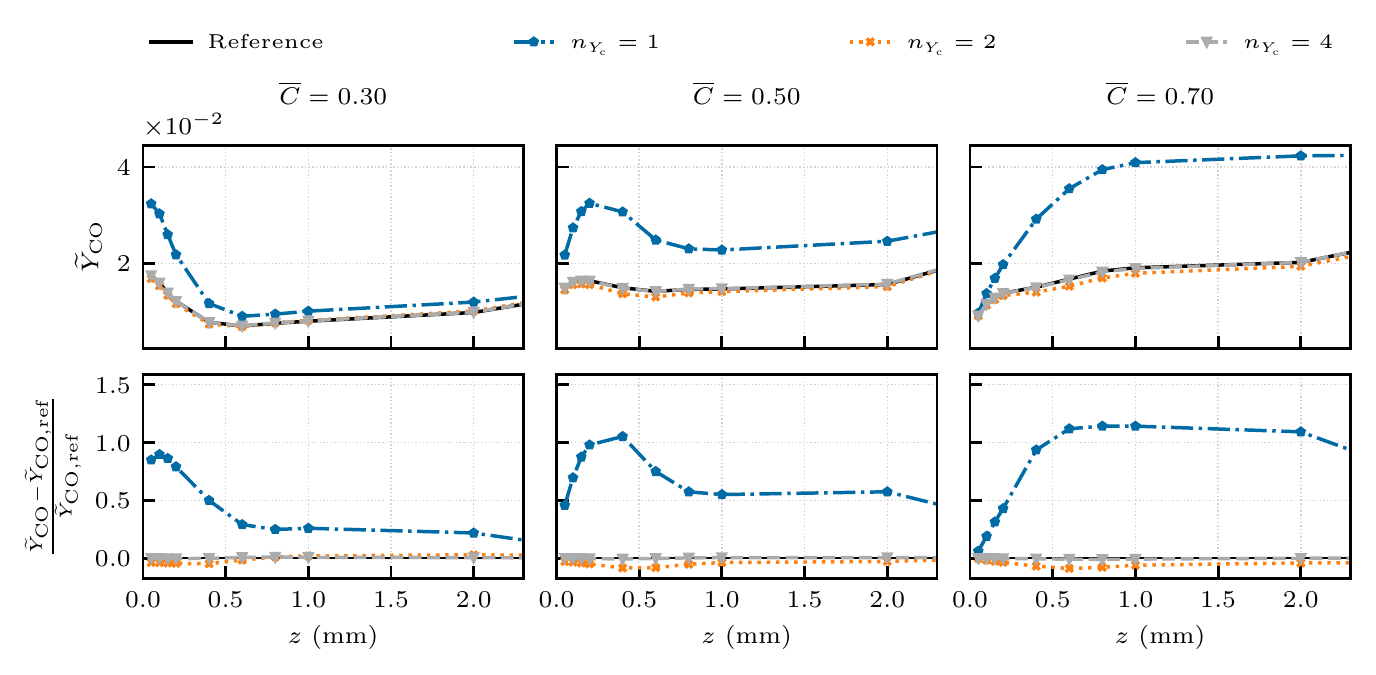}
  \caption{Prediction of the Favre-averaged CO mass fraction $\widetilde{Y}_\text{CO}$ at different wall-normal positions $z$ in the context of RANS. The CQMOM approximation is performed using a different number of nodes in the progress variable direction $n_{Y_\mathrm{c}}$, while in the enthalpy direction only a single node is used ($n_{H}=1$). The different columns show different stream-wise locations in the flame defined by $\overline{C}$ at the respective wall distance, see Section~\ref{sec:PDFs}. At the top the absolute value is shown, while at the bottom the relative deviation from the quasi-DNS reference is depicted.}
  \label{fig:RANS-convergence-CO}
\end{figure}

Figure~\ref{fig:RANS-convergence-CO} shows the convergence for the Favre-averaged CO mass fraction for an increasing number of nodes in the direction of the progress variable $n_{Y_\mathrm{c}}$ and with one node in the direction of the enthalpy ($n_{H}=1$). When only a single node in the progress variable direction is employed, the CQMOM approach shows a high model deviation compared to the reference DNS results.
In this particular case, only the Favre-averaged mean of the progress variable $\widetilde{Y}_\mathrm{c}$ and normalized enthalpy $\widetilde{H}$ are used for the determination of the Favre-averaged CO mass fraction. Using two integration nodes, the CQMOM approach already shows a high prediction accuracy both in the core flow and in the proximity to the wall. With four integration nodes in the progress variable direction and a single node in the enthalpy direction, a nearly perfect agreement between the CQMOM estimate and the quasi-DNS reference can be observed. For the context of LES a corresponding study was performed, using the respective spatially filtered values, leading to similar results that are not shown here for brevity. In the following analysis, all RANS and LES CQMOM results are shown for four nodes in the progress variable direction and a single node in the enthalpy direction.

\subsection{\textit{A priori} assessment of the CQMOM approach in the context of RANS}
\label{sec:CQMOM-TCI-RANS}
\begin{figure}[htb]
  \includegraphics[scale=1.0]{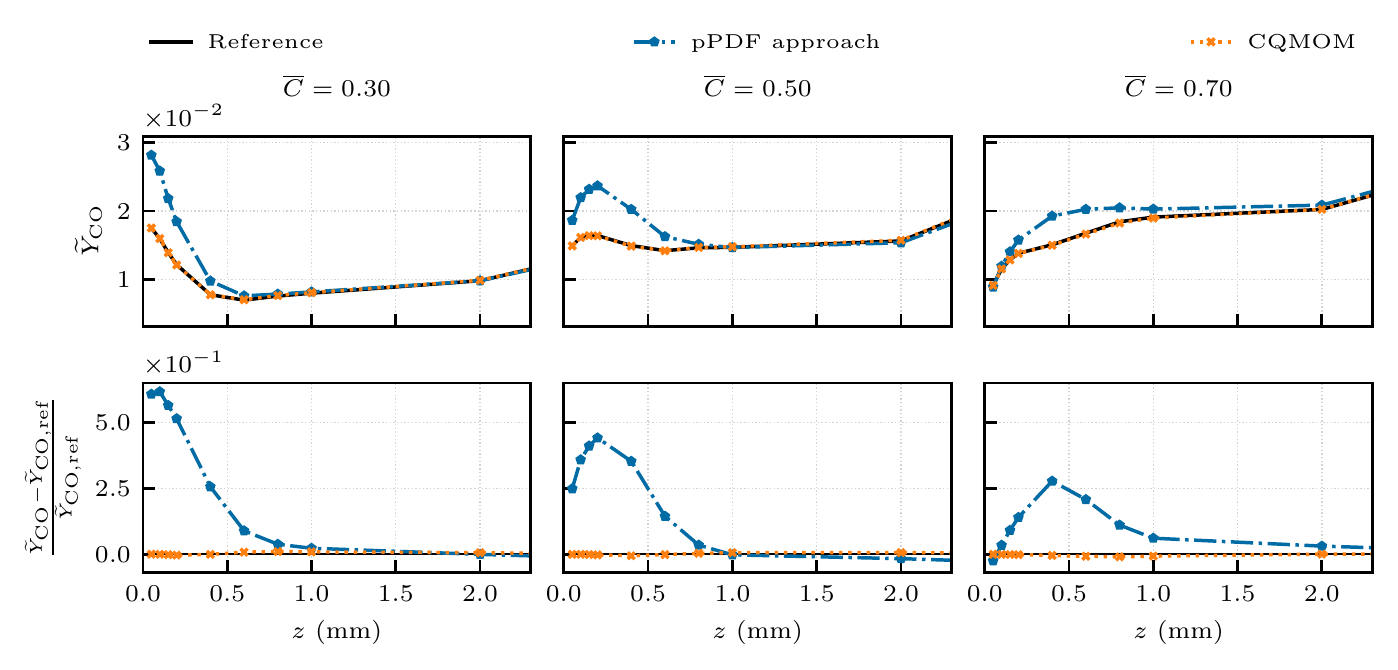}
  \caption{Prediction of the Favre-averaged CO mass fraction $\widetilde{Y}_\text{CO}$ at different wall distances $z$ in the context of RANS. Next to the CQMOM a pPDF approach from the literature~\cite{Fiorina2005} is shown. The different columns show different stream-wise locations in the flame defined by $\overline{C}$ at the respective wall distance, see Section~\ref{sec:PDFs}. At the top the absolute value is shown, while at the bottom the relative deviation from the quasi-DNS reference is depicted.}
  \label{fig:RANS-QMOM-lookup-CO}
\end{figure}

\begin{figure}[htb]
  \includegraphics[scale=1.0]{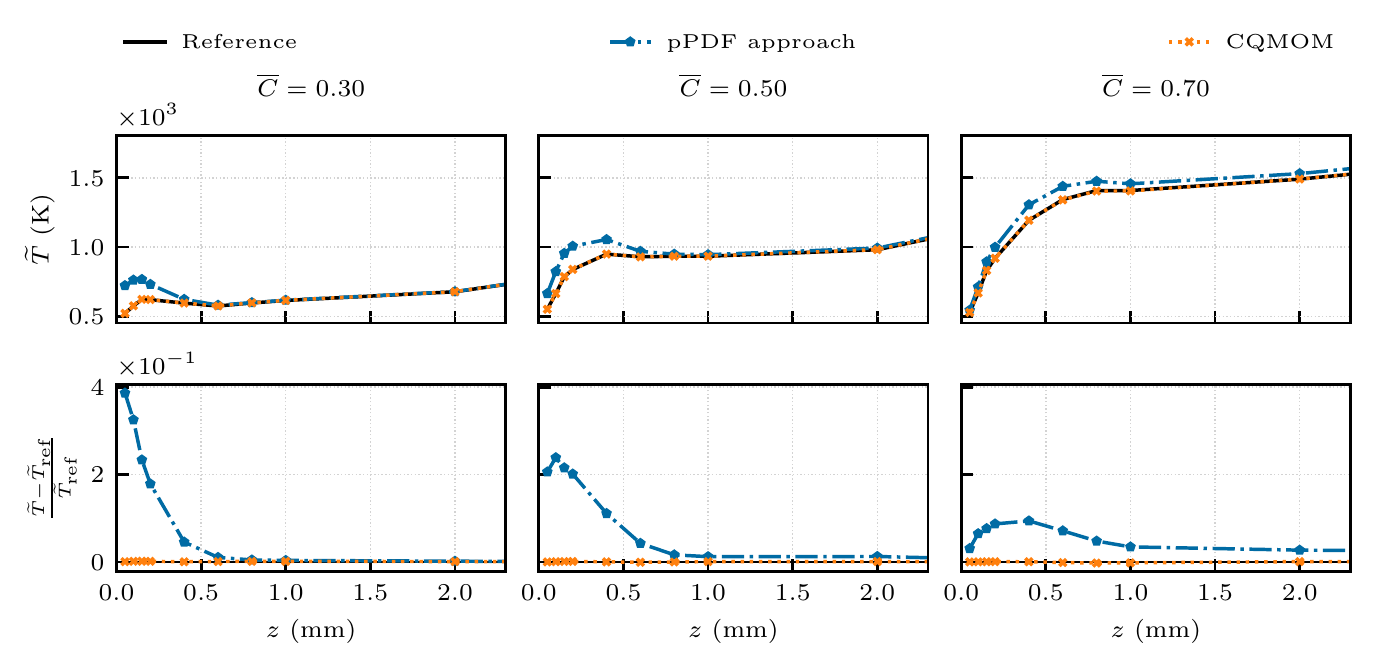}
  \caption{Prediction of the Favre-averaged temperature $\widetilde{T}$ as a function of wall distances $z$ in the context of RANS. Next to the CQMOM a pPDF approach from the literature~\cite{Fiorina2005} is shown. The different columns show different stream-wise locations in the flame defined by $\overline{C}$ at the respective wall distance, see Section~\ref{sec:PDFs}. At the top the absolute value is shown, while at the bottom the relative deviation from the quasi-DNS reference is depicted.}
  \label{fig:RANS-QMOM-lookup-T}
\end{figure}

Figures~\ref{fig:RANS-QMOM-lookup-CO} and~\ref{fig:RANS-QMOM-lookup-T} show the CQMOM prediction of the Favre-averaged CO mass fraction and temperature in comparison to the pPDF approach from the literature~\cite{Fiorina2005}.
At the top row, the Favre-averaged quantity is shown, while the relative deviation from the quasi-DNS reference is depicted at the bottom.

Different stream-wise locations in the reaction zone of the flame are displayed, with increasing reaction progress from left to right.
The free flow region ($z>1~\mathrm{mm}$) that is unaffected by enthalpy losses to the wall confirms that in this region both models are in excellent agreement with the reference, both for the CO mass fraction and the temperature. Due to negligible enthalpy losses, the PDFs reduce to a univariate distribution solely dependent on the progress variable, see Section~\ref{sec:PDFs-RANS}.
With decreasing distance to the wall ($z<1~\mathrm{mm}$), the flame is affected by increasing enthalpy losses leading to a bivariate PDF of the progress variable and enthalpy. In this close proximity to the wall, the prediction accuracy of the pPDF approach deteriorates substantially with deviations up to 50\% from the quasi-DNS reference. Here, the pPDF approach suffers from two drawbacks: (i) the enthalpy PDF cannot be modeled by a simple $\delta$-peak and (ii) the modeling assumption of independence of the normalized progress variable and normalized enthalpy~\cite{Fiorina2005} is no longer valid in the close vicinity of the wall. Contrary, the CQMOM approach can capture the near-wall behavior accurately by accounting for the conditional PDF of the enthalpy for a given progress variable $P \left( Y_\mathrm{c} | H \right)$ using only a single node in the enthalpy direction. In the close vicinity to the wall the CQMOM approach makes use of the additional moment information compared to the pPDF approach and is able to approximate the complex bivariate PDF of the progress variable and enthalpy leading to an excellent agreement with the quasi-DNS reference.

\newpage

\subsection{\textit{A priori} assessment of the CQMOM approach in the context of LES}
\label{sec:CQMOM-TCI-LES}
Finally, the suitability of the CQMOM approach to model the different FDF shapes presented in Section~\ref{sec:PDFs-LES} is discussed. The prediction accuracy of the Favre-filtered CO mass fraction is assessed, since it showed the highest deviations in the context of RANS and, hence, is most challenging to model. Two scenarios are studied, the flame flank (A) and the flame tip (B) for different wall distances $z$ and
wall-normal filter width $\delta_z$. Furthermore, the influence of the filter kernel's aspect ratio $\mathrm{AR}$ is discussed for $\delta_z = 0.2~\mathrm{mm}$. The stream-wise location of the box filer's center point corresponds to a value of $C=0.5$.

\begin{figure}[htb]
  \includegraphics[scale=1.0]{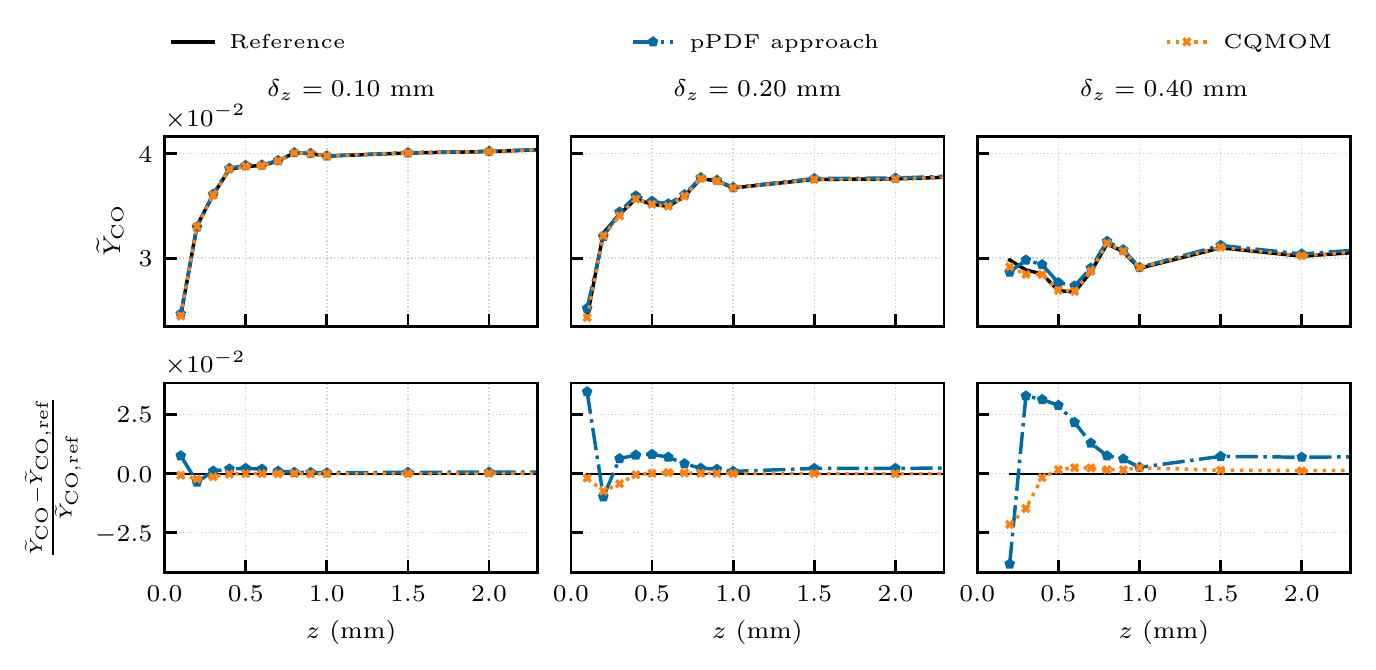}
  \caption{Prediction of the instantaneous Favre-filtered CO mass fraction as a function of wall distance $z$ in the context of LES. The data is extracted at the flame flank (A) at a stream-wise direction corresponding to $C=0.5$, see Section~\ref{sec:PDFs-LES}. The different columns show different box filter width, see Eqn.~\eqref{eq:box-filter-size}, with varying wall-normal filter width $\delta_z$ and $\mathrm{AR}=2.5$. At the top the absolute value is shown, while at the bottom the relative deviation from the quasi-DNS reference is depicted.}
  \label{fig:LES-QMOM-lookup-CO-A}
\end{figure}

\begin{figure}[htb]
  \includegraphics[scale=1.0]{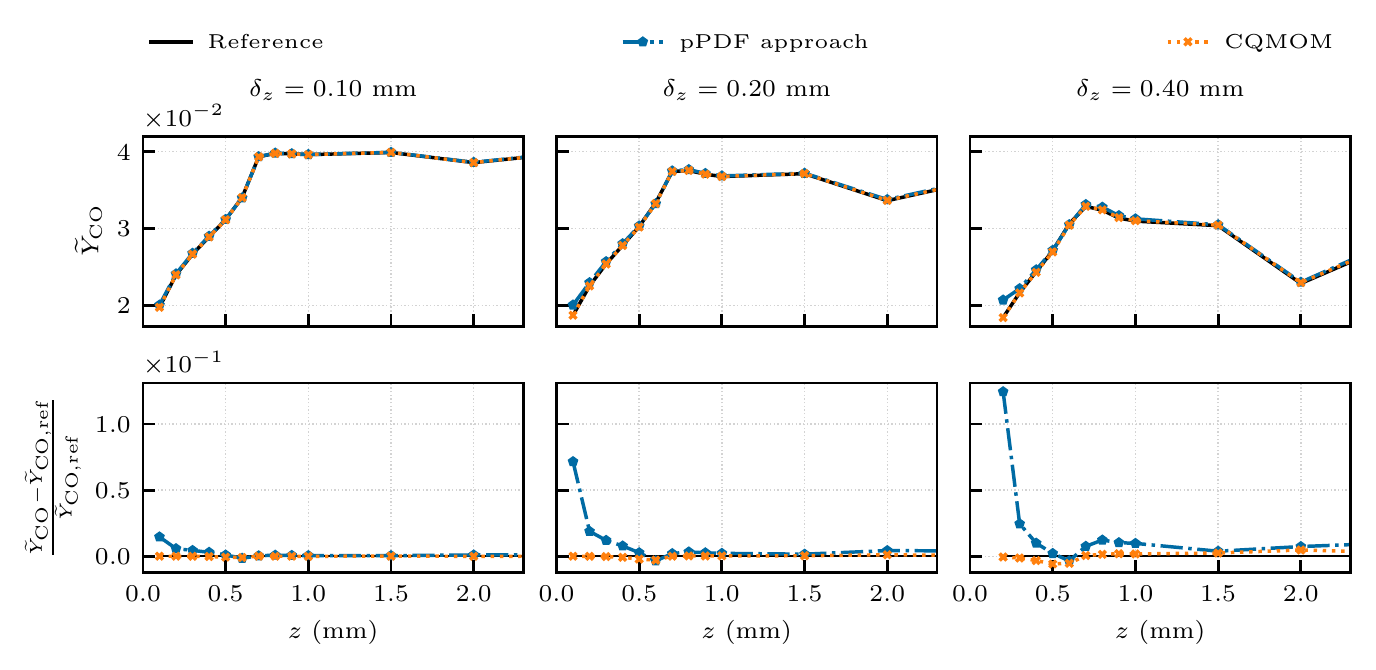}
  \caption{Prediction of the instantaneous Favre-filtered CO mass fraction as a function of wall distance $z$ in the context of LES. The data is extracted at the flame tip (B) at a stream-wise direction corresponding to $C=0.5$, see Section~\ref{sec:PDFs-LES}. The different columns show different box filter width, see Eqn.~\eqref{eq:box-filter-size}, with varying wall-normal filter width $\delta_z$ and $\mathrm{AR}=2.5$. At the top the absolute value is shown, while at the bottom the relative deviation from the quasi-DNS reference is depicted.}
  \label{fig:LES-QMOM-lookup-CO-B}
\end{figure}

Figures~\ref{fig:LES-QMOM-lookup-CO-A} and~\ref{fig:LES-QMOM-lookup-CO-B} show the results for varying wall-normal filter width and an aspect ratio of $\mathrm{AR}=2.5$ for the flame flank and the flame tip, respectively. Again, both closure models show a high prediction accuracy in the free flow ($z>1.0~\mathrm{mm}$). However,  closer to the wall ($z<0.5~\mathrm{mm}$) the prediction accuracy deteriorates particularly for the pPDF approach. At the flame flank in Fig.~\ref{fig:LES-QMOM-lookup-CO-A}, the CQMOM predictions show a relative difference of up to 2.5\% from the DNS reference for the highest wall-normal filter width $\delta_z=0.4~\mathrm{mm}$, while still remaining a higher prediction accuracy compared to the pPDF approach. With decreasing filter width the prediction of the PDF models improves significantly. As discussed in the previous section, the flame flank has a wide FDF of the progress variable and enthalpy and, therefore, it is particularly challenging to model with the CQMOM approach. This can be overcome in two ways: (i) the number of nodes in the enthalpy direction can be increased and (ii) the filter width at the wall can be decreased (e.g. the grid resolution is increased), leading to less complex FDFs. 
At the flame tip in Fig.~\ref{fig:LES-QMOM-lookup-CO-B} the progress variable and enthalpy show a very high correlation. In this area, the CQMOM modeling assumption of a conditional FDF of enthalpy for a given progress variable is especially advantageous. Here, the CQMOM shows a nearly perfect agreement with the reference. The pPDF approach, on the other hand, assuming the independence of progress variable and enthalpy is not able to predict the correct Favre-filtered CO mass fraction in the close vicinity of the wall.

\begin{figure}[htb]
  \includegraphics[scale=1.0]{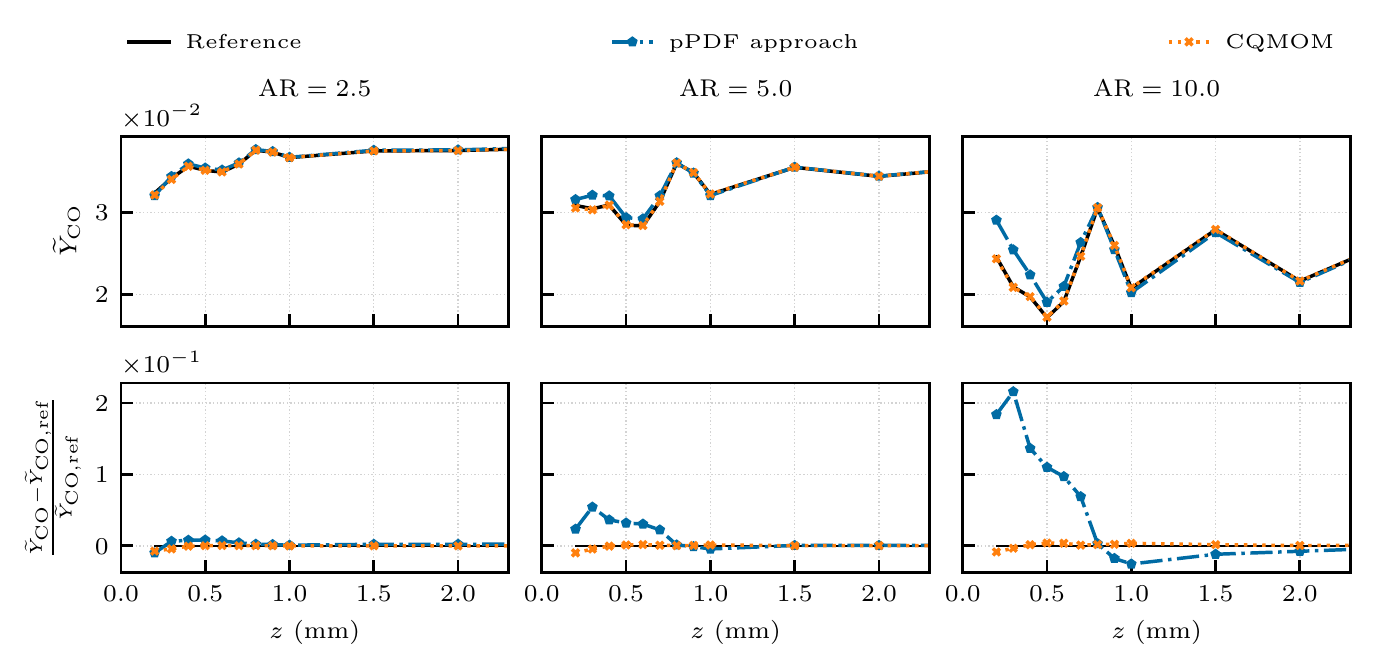}
  \caption{Prediction of the instantaneous Favre-filtered CO mass fraction as a function of wall distance $z$ in the context of LES. The data is extracted at the flame flank (A) at a stream-wise direction corresponding to $C=0.5$, see Section~\ref{sec:PDFs-LES}. The different columns show different box filter aspect ratios, see Eqn.~\eqref{eq:box-filter-size}, for $\delta_z=0.2~\mathrm{mm}$. At the top the absolute value is shown, while at the bottom the relative deviation from the quasi-DNS reference is depicted.}
  \label{fig:LES-QMOM-lookup-CO-A-AR}
\end{figure}

Finally, the influence of the box filter's aspect ratio is discussed. As discussed in section~\ref{sec:PDFs-LES}, the FDF shape at the flame flank is significantly influenced by the aspect ratio. In Fig.~\ref{fig:LES-QMOM-lookup-CO-A-AR} the results for the flame flank are shown. While the prediction accuracy of the pPDF approach deteriorates with increasing aspect ratio, the CQMOM is able to correctly model the FDF showing a high accuracy for all aspect ratios. This emphasises the importance of a suitable FDF closure model with decreasing LES resolution. For the flame tip the aspect ratio has only a small influence on the prediction accuracy of both PDF closure models. The corresponding results are therefore not shown.

\newpage

\section{Conclusion}
Turbulent FWI in a generic turbulent side-wall quenching configuration is investigated in this work. A quasi-DNS of a stoichiometric methane-air flame was performed. From the quasi-DNS data the joint PDFs/FDFs of the progress variable and normalized enthalpy were extracted and analyzed in the context of RANS/LES.
The influence of enthalpy losses at the wall on the joint PDFs/FDFs was assessed for different wall distances and mean reaction progress in the flame. Analyzing the PDFs in the context of RANS clearly showed that the wall distance has a strong influence on the PDF shape and dimension. While in the core flow a univariate PDF solely dependent on the progress variable is observed, the PDF shape becomes increasingly complex and bivariate closer to the wall. In the context of LES, the FDF has a high dependency on the spatial position in the flow and two representative flame positions were discussed: a flame flank (A) and a flame tip (B). At the flame flank a wide FDF was present that showed a high dependency on the box filter's aspect ratio and size (e.g. LES grid resolution). At the flame tip the FDF is less affected by the box filter's shape, however, a strong correlation was observed between the progress variable and normalized enthalpy close to the wall. This finding is in contradiction to the often used modeling assumption of statistical independence, i.e. in pPDF approaches~\cite{Fiorina2005, Donini2017, Zhang2021}.

In the second part of this work, a novel CQMOM approach, coupled to a Quenching-Flamelet Generated manifold is assessed in an \textit{a priori} analysis in the context of RANS and LES. The results are compared to the quasi-DNS data and a pPDF approach from the literature using a $\beta$-PDF for the progress variable and a $\delta$-peak for enthalpy. First, a moment convergence study was performed, showing very good agreement with the quasi-DNS reference for four nodes in the progress variable direction and a single node in the enthalpy direction. In the core flow region unaffected by wall heat losses, both models are in good agreement with the quasi-DNS reference. In close proximity to the wall, however, the pPDF approach is not able to capture the correct flame behavior in the context of RANS and for coarse LES (large box filter sizes). The CQMOM approach, on the other hand, shows excellent agreement with the quasi-DNS reference both in the near-wall region and in the core flow. These results are very promising for future fully coupled LES or RANS simulations and provide an alternative approach to pPDF to account for TCI.


\section*{Acknowledgments}
This work has been funded by the Deutsche Forschungsgemeinschaft (DFG, German Research Foundation)—Projektnummer 237267381—TRR 150. Calculations for this research were conducted on the Lichtenberg high performance computer of the TU Darmstadt. The DNS simulation was performed on the national supercomputer HAWK at the High Performance Computing Center Stuttgart (HLRS) under the grant number DNSbomb/xbifezh and on the HoreKa supercomputer funded by the Ministry of Science, Research and the Arts Baden-Württemberg and by the Federal Ministry of Education and Research. Finally, we would like to thank Dr. Martin Pollack for his contribution and discussion about QbMM closure.

\section*{Data availability statement}
The data presented in this study is available as supplementary material under \url{https://doi.org/10.48328/tudatalib-673}.

\bibliography{publication.bib}
\bibliographystyle{unsrtnat_mod}

\newpage

\appendix
\section{Favre-averaging in the context of RANS and LES}
\label{appendix:averaging}
To calculate a Favre-averaged quantity in the context of RANS a temporal average is performed and the Favre-averaged quantity $\widetilde{Q}\left( \mathbf{x}, t \right)$ is given by
\begin{align}
  \widetilde{Q}_\text{RANS}\left( \mathbf{x} \right) = \frac{\int_{-T/2}^{T/2} \rho \left(  \mathbf{x}, t \right) Q \left( \mathbf{x}, t \right)  dt}{\int_{-T/2}^{T/2} \rho \left( \mathbf{x}, t \right) dt} \ ,
\end{align}

\noindent where $t$ corresponds to the time, $T$ is the temporal integration interval and $\rho$ the density. In the investigated channel flow in this work, the averaging is additionally performed in the statistical independent lateral direction $y$, leading to

\begin{align}
 \widetilde{Q}_\text{RANS}\left( \mathbf{x} \right) = \frac{\int_{y_\mathrm{min}}^{y_\mathrm{max}}\int_{-T/2}^{T/2} \rho \left(  \mathbf{x}, t \right) Q \left( \mathbf{x}, t \right)  dtdy}{\int_{y_\mathrm{min}}^{y_\mathrm{max}}\int_{-T/2}^{T/2} \rho \left( \mathbf{x}, t \right) dtdy} \ ,
\end{align}
\noindent with $y_\mathrm{min}$ and $y_\mathrm{max}$ being the minimum and maximum lateral coordinate, respectively.
In LES a spatial filter operation is performed and $\widetilde{Q}\left( \mathbf{x}, t \right)$ is given by

\begin{align}
  \widetilde{Q}_\text{LES} \left( \mathbf{x}, t \right) = \frac{\int \rho \left( \mathbf{x}^*, t \right) Q \left( \mathbf{x}^*, t \right) F \left( \mathbf{x} - \mathbf{x}^* \right) d\mathbf{x}^* }{\int \rho \left( \mathbf{x}^*, t \right) F \left( \mathbf{x} - \mathbf{x}^* \right) d \mathbf{x}^*} \ ,
\end{align}

\noindent where $\mathbf{x}=\left( x_1, x_2, x_3 \right)$ is the spatial coordinate and $F \left( \mathbf{x} - \mathbf{x}^* \right)$ is a normalized spatial filter function. In the following analysis a box filter is applied given by

\begin{align}
  \label{eq:box-filter}
  F \left( x_{1}, x_{2}, x_{3} \right)= \left \{
   \begin{array}{ll}
       \left( \Delta_{x_1} \cdot \Delta_{x_2} \cdot \Delta_{x_3} \right)^{-1} & \text { if } \left| x_{i} \right| \leq \Delta_{x_i} / 2, i=1,2,3 \\
      0 & \text { otherwise }
  \end{array}
  \right. \ .
\end{align}

\section{QFM validation in turbulent side-wall quenching}
\label{sec:QFM-validation}
\begin{figure}[htb]
    \centering
  \includegraphics[scale=1.0]{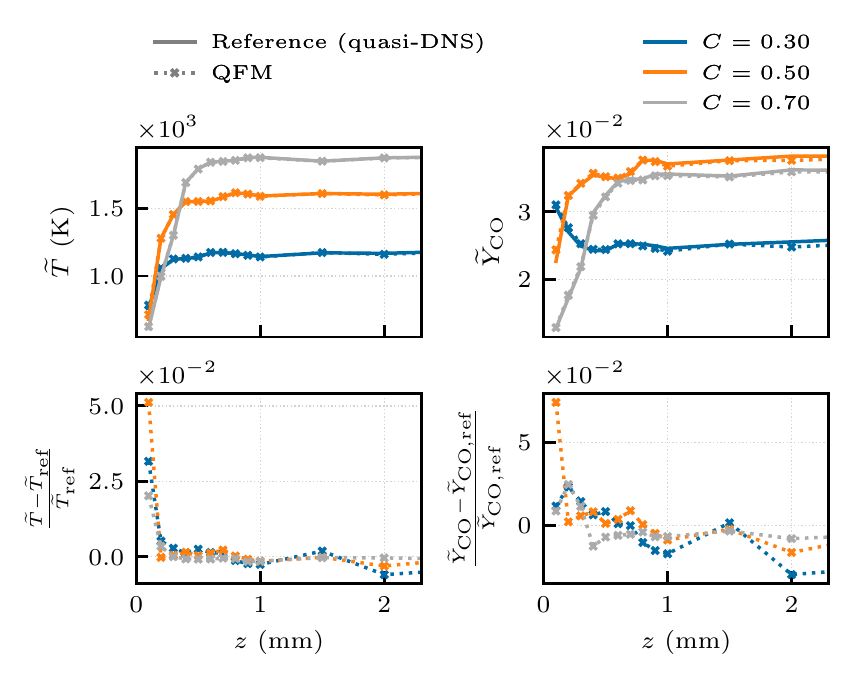}
  \caption{Instantaneous Favre-filtered temperature (left) and CO mass fraction (right) calculated directly from the quasi-DNS and using a lookup with the QFM (top) and the relative error of the means (bottom). The plot shown corresponds to the flame flank (A), see Fig.~\ref{fig:LES-sampling}.}
  \label{fig:LES-QFM-validation-A}
\end{figure}

\begin{figure}[htb]
    \centering
   \includegraphics[scale=1.0]{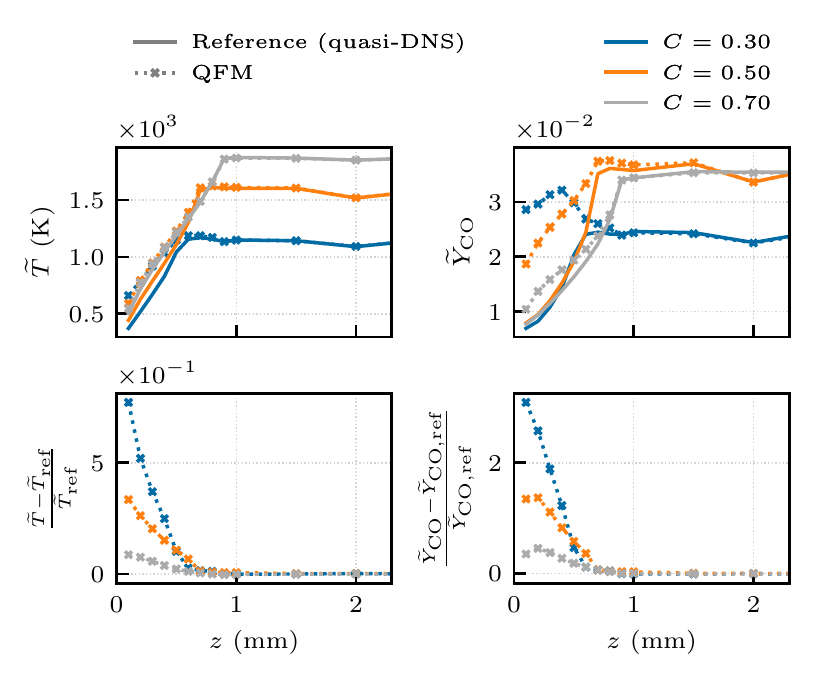}
   \caption{Instantaneous Favre-filtered temperature (left) and CO mass fraction (right) calculated directly from the quasi-DNS and using a lookup with the QFM (top) and the relative error of the means (bottom). The plot shown corresponds to the flame tip (B), see Fig.~\ref{fig:LES-sampling}.}
  \label{fig:LES-QFM-validation-B}
\end{figure}

In this section, the QFM manifold is validated for the turbulent side-wall quenching. Therefore, Figs.~\ref{fig:LES-QFM-validation-A} and~\ref{fig:LES-QFM-validation-B} show the Favre-filtered temperature and CO mass fraction calculated directly from the quasi-DNS and estimated with the QFM through a table lookup.
The data is extracted with a wall normal filter width of $\delta_z = 0.2~\mathrm{mm}$ and $\mathrm{AR}=2.5$ at different stream-wise locations defined by the normalized progress variable $C$, see Eqn.~\eqref{eq:C}.
The quasi-DNS reference quantity $\widetilde{Q}$ is calculated as

\begin{equation}
  \widetilde{Q} = \frac{\sum_{i=0}^{N} \rho_{i} Q_{i}}{ \sum_{i=0}^{N} \rho_{i} } \ ,
\end{equation}

\noindent while, for the QFM value, both the density and the quantity are estimated from the chemistry manifold:

\begin{equation}
  \widetilde{Q}_\text{QFM} = \frac{\sum_{i=0}^{N} f_\rho \left( C_{i}, H_{i} \right) f_Q \left( C_{i}, H_{i} \right) } { \sum_{i=0}^{N} f_\rho \left( C_{i}, H_{i} \right) } \ .
\end{equation}

\noindent While the manifold shows very good agreement with the quasi-DNS in the free flow and at the flame flank (A), at the flame tip (B) the manifold shows a deviation from the quasi-DNS reference in the near-wall region. These deviations in the manifold are due to turbulent mixing processes that do not occur in the laminar flames and are subject to future investigations.

\end{document}